\documentclass[%
 preprint,
 amsmath,amssymb,
 aps, physrev,
]{revtex4-2}

\usepackage{graphicx}% Include figure files
\usepackage{dcolumn}% Align table columns on decimal point
\usepackage{bm}% bold math
\usepackage[colorlinks=true,urlcolor=blue,linkcolor=blue,citecolor=blue]{hyperref}
\usepackage{subfigure}
\usepackage{multirow}
\usepackage{array}
\usepackage{caption}
\begin{document}

\preprint{APS/123-QED}

\title{\textbf{Effect of correlation on the elastic scattering of slow positrons from molecules.} 
}% 

\author{Snigdha Sharma}
 % \altaffiliation[Also at ]{Department of Physics,School of Advanced Sciences, Vellore Institute of Technology,\\ Vellore, Tamil Nadu, India - 632014.}%Lines break automatically or can be forced with \\
\author{Dhanoj Gupta}%
 \email{Contact author: dhanojsanjay@gmail.com}
\affiliation{%
 % Authors' affiliations\\
  Department of Physics, School of Advanced Sciences, Vellore Institute of Technology,\\ Vellore, Tamil Nadu, India - 632014.
}%

%  \homepage{http://www.Second.institution.edu/~Charlie.Author}
% \affiliation{
%  First affiliation for this author
% }%
% \affiliation{
%  second institution for this author
% \affiliation{%
%  Authors' institution and/or address\\
%  This line break forced with \textbackslash\textbackslash
% }%

% \collaboration{CLEO Collaboration}%\noaffiliation

\date{\today}% It is always \today, today,

\begin{abstract}
 Many-body correlation plays a crucial role in the low-energy positron-molecule scattering dynamics. In the present work, we have integrated a recent model correlation potential, developed by Swann and Gribakin, with the single-center expansion method and verified its efficacy in reproducing the best of \textit{ab initio} and experimental results, for both non-polar and polar molecules. Starting with the already tested molecules---hydrogen, ethylene, and acetylene---with the model correlation, we extended our calculations to oxygen-containing molecules: oxygen, water, and formic acid. In order to provide a comparative study, we have also performed calculations employing the model correlation of Perdew and Zunger. Integral, differential, and momentum transfer cross-sections are reported for the target molecules. Positron virtual/bound state formation is also predicted using both of the model correlations. Overall, an improved agreement of our results with the literature for the recent model suggests the approach can be employed for larger systems, where the \textit{ab initio} techniques are difficult to implement.      
% \begin{description}
% Secondary publications and information retrieval purposes.
% You may use the \texttt{description} environment to structure your abstract;
% use the optional argument of the \verb+\item+ command to give the category of each item. 
% \end{description}
\end{abstract}

\maketitle

%\tableofcontents

\section{\label{sec:level1}Introduction\protect}
% The line
% break was forced \lowercase{via} \textbackslash\textbackslash}
Positron interaction with molecules entails intricate correlations, which widely affect the scattering dynamics, typically in the low-energy regime. Theoretical investigation of such interactions involves either an \textit{ab initio} description of the positron-molecule correlation-polarization potential or a model description of the same. The former, though more accurate, is limited to scattering studies that involve smaller molecules. As the size of the target increases, the calculations become computationally expensive, and the accurate description of the correlation-polarization potential becomes challenging \cite{Review_exp_theory_2017, frighetto2024low}. R-matrix \cite{R-matrix}, Schwinger multichannel (SMC) \cite{SMC}, and convergent close coupling (CCC) \cite{CCC} are a few widely used \textit{ab initio} methods for positron-molecule scattering calculations.  

Another aspect of positron-target interaction that needs substantial improvement, both experimentally and theoretically, is the binding of positrons with atoms and molecules. Positron approaching the target can undergo a direct annihilation with the target electron or attach itself to the target, forming a bound state. Positron-molecule correlations, such as polarization due to the incoming positron, screening of positron-electron Coulomb interaction, and virtual-positronium formation (temporary tunneling of molecular electron to the positron), greatly influence the binding of positron to targets \cite{Gribakin,hofierka2024many,hofierka2022many}. The initial theoretical prediction of positron bound states with Mg, Zn, Cd, and Hg atoms was suggested by Dzuba \textit{et al.} \cite{Dzuba} using the many-body perturbation theory (MBPT). Since then, a number of atomic targets have been shown to support positron bound states \cite{Mitroy_2002,Cheng,Dzuba_2012,Bubin,Harabati,Brorsen,Amaral}. There hasn't been any experimental detection of positron-atom bound states due to limitations in obtaining atoms in the gaseous phase \cite{Tumakov,swann2021effect}. However, in the case of molecular targets, positron bound states have been detected experimentally for nearly 90 molecules, mostly nonpolar or weakly polar, by studying the energy resolved annihilation spectra \cite{hofierka2024many}. In the theoretical front, \textit{ab initio} calculations of positron binding energy, using approaches like configuration interaction (CI) \cite{Chojnacki10072006, Tachikawa_2003, Tachikawa_2012} and the any particle molecular orbital (APMO) \cite{romero, APMO} have been reported, in particular, for polar molecules. However, the results vary significantly from the experimental predictions as these methods rely on the single-particle basis center for the expansion of positron wave function, which is insufficient to completely describe the positron-target interaction \cite{Charry,hofierka2022many}. Hofierka \textit{et al.} developed a many-body theory for positron interaction with polyatomic molecules that quantitatively accounts for the many-body correlations, thereby providing good agreement of the calculated binding energies with experimental data \cite{hofierka2022many,hofierka2024many}. Furthermore, a machine-learning method was adopted by Amaral \textit{et al.} \cite{amaral2020machine} to predict the binding energies for a set of polar and non-polar molecules. 

To circumvent the limitations and complications in \textit{ab initio} techniques, a model potential approach was developed by Swann and Gribakin \cite{Swann2018} to calculate the positron binding energy for both polar and non-polar molecules. The subsequent calculations using the model potential (which takes care of the virtual positronium formation), for atoms and molecules, including hydrocarbons, gave an overall good result in accord with the experiments \cite{Swann2019,swann2020model,swann2021effect}. The model potential of Swann and Gribakin was integrated with the SMC method by Frighetto \textit{et al.} \cite{Frighetto2023Imp} to replace the prevailing static-polarization (SP) approximation in \textit{ab initio} methods, wherein the polarization of the target is approximated as the single virtual excitations of the target electrons. The approximation demands the use of extra chargeless centres and diffuse functions to improve the description of the positron-target interaction, amplifying the computational cost and limiting the size of the target that can be studied. The use of model potential in the SMC method not only enabled to predict valid bound states in molecules but also improved the results of scattering observables (integral and differential cross-sections). This was testified in Frighetto's recent works \cite{Frighetto2023Imp, frighetto2023low, frighetto2024low}.                                      

In a nutshell, a reliable model of the correlation-polarization potential between the incident positron and the electrons of the target molecule would be a boon in maintaining the accuracy and cost-effectiveness of the scattering calculations, even with larger targets. 

In this work, we have integrated the model potential by Swann and Gribakin \cite{Swann2018} with the single-center expansion (SCE) method. The method, already validated to work with larger molecules for positron scattering, like DNA nucleobases and their analogues \cite{franz2014low, franz2013low, franz2013low1,Sharma_2025}, forms a standard framework to testify the applicability and the accuracy of new model potentials to calculate the scattering observables as well as the positron binding energies. The aim of this work is to initially reproduce the integral cross-sections calculated by Frighetto \textit{et al.} for hydrogen molecule, ethylene, and acetylene utilizing the model potential of Swann and Gribakin. From this, we will estimate the cut-off radius (refer to section \ref{subsec:level4A}) for the constituent atoms, a free parameter in the model correlation. We will then extend our work to include the scattering calculations for oxygen molecule and water, providing us the cut-off radius for oxygen atom. 

Making use of the cut-off radii of hydrogen, carbon, and oxygen atoms, we will calculate the integral (ICS), differential (DCS), and momentum transfer (MTCS) cross-sections for formic acid. Formic acid is chosen for the current study due to the disagreements between the available theoretical data and experimental cross-sections \cite{stevens} in the literature. Also, this work is an opportunity to demonstrate the applicability of the model potential to oxygen-containing molecules. Due to some discrepancies and suggestions for more theoretical studies---as highlighted in the literature \cite{Tenfen_2019,Loreti_2016}---on low-energy cross sections, specifically DCS, we report the ICS, DCS, and MTCS for oxygen molecule and water as well. All the aforementioned calculations are repeated using the correlation potential of  Perdew and Zunger \cite{Perdew}, in order to give a clear comparison. Finally, the positron virtual/binding energies for all the targets are also evaluated and compared with the available data in the literature, providing an overall picture of the role played by positron-molecule correlation on the various aspects of low-energy interaction.  

The rest of the paper is arranged as follows: section \ref{sec:level2} describes the methodology employed, section \ref{sec:level3} is the computational details, in section \ref{sec:level4} we have the results and discussion, and section \ref{sec:level5} is the conclusion.

\section{\label{sec:level2}Methodology\protect}
\subsection{Theoretical background}
The theoretical basis for this work is the single-center expansion (SCE) method, applied here for elastic scattering calculations. The method adheres to the fixed-nuclei (FN) approximation, and the target molecule remains fixed in its ground-state geometry. For computational ease, the scattering dynamics are treated in the body-fixed frame. While the essential aspects of the SCE formalism are outlined here, readers are directed to the literature for a more detailed explanation \cite{BACCARELLI20111,Winifred}. In this approach, the potentials, bound wavefunctions, and continuum wavefunctions are expanded in terms of symmetry-adapted angular functions centered at a common origin, typically the center of mass of the target molecule. 
 
The bound wavefunction, $\phi_i$, for a single electron is expressed as, 

\begin{align}
 \mathcal \phi_{i}^{p\mu} (\textbf{r})= \frac{1}{r} \left[\sum_{hl} u_{hl}^{i,p\mu} (r) X_{hl}^{p\mu} (\theta,\phi)\right].
\label{eqn1}
\end{align}
Here, index $i$ labels a specific multicentre bound orbital belonging to a particular irreducible representation (irrep) $p$ of the molecular point group. $\mu$ is one of the components of the irrep. The index $h$ denotes a basis function for a given partial wave $l$.  

The spherical harmonics \textcolor{black}{(real spherical harmonics in case of closed-shell molecules)} are combined linearly to form a symmetry-adapted angular function,

\begin{align}
 X_{hl}^{p\mu} (\theta,\phi)=\sum_{m=-l}^{l} b_{hlm}^{p\mu}Y_{l}^{m} (\theta,\phi). 
\end{align}
The expansion coefficients $b$ can be determined using the character tables for each irrep of the molecular point group. 
% Likewise, the scattering electron wavefunction and the potentials describing the electron-molecule interaction are expanded. 
The expansion of Eq.\ref{eqn1} requires a multicentre wavefunction of the target molecule, which can be obtained from quantum chemistry software, such as Gaussian \cite{g16}. The method of angular quadrature is then used to determine the radial coefficients, such as the following for multicentre Gaussian-type orbitals (GTOs),

 % \begin{figure*}
\begin{align}
\begin{split}
  u_{hl}^{i}(r;R) &= \sum_{k,j,\nu,m} \int_{0}^{\pi}sin(\theta)d\theta \int_{0}^{2\pi} b_{hlm}^{i}Y_{l}^{m} (\theta,\phi)\\
  & C_{kj}^{i}(R)d_{\nu}^{kj}g_{\nu}^{kj}(x_{k})d\phi,
\label{eqn3}
\end{split}
\end{align}
 % \end{figure*}
where $C_{kj}^{i}$ is the coefficient of $\mu^{th}$ GTO for $k^{th}$ atomic centre at a given molecular geometry R. $d_{\nu}^{kj}$ denotes the contraction coefficients of the cartesian gaussian functions $g_{\nu}^{kj}$. 

Once the radial coefficients are determined, the bound wavefunction is used to compute the electron density, which in turn allows for the calculation of the electrostatic potential and the correlation-polarization potential. Eventually, we get a set of coupled radial equations for the scattering process,

\begin{align}
 \left[\frac{d^2}{dr^2}-\frac{l(l+1)}{r^2}+k^2\right] F_{lh}^{p\mu}(r)= 2\sum_{l'h'}V_{lh,l'h'}^{p\mu}(r)F_{l'h'}^{p\mu}(r),
 \label{eqn4}
\end{align}
which are solved to get the scattering parameters, viz, K- and T-matrix elements in the body-fixed frame. Here, $V_{lh,l'h'}^{p\mu}(r)$ are the coupling potential elements given as,

\begin{align}
 V_{lh,l'h'}^{p\mu}(r) = \int X_{hl}^{p\mu}(\theta,\phi)V(\textbf{r})X_{h'l'}^{p\mu}(\theta,\phi)d\theta d\phi  
\end{align}
\begin{align}
and, V(\textbf{r}) = V_{st} + V_{corr,pol} \nonumber
\end{align}
The static potential ($V_{st}$) is calculated exactly using the molecular electron density $\rho(\textbf{r})$, whereas a local model is considered for the correlation-polarization ($V_{corr,pol}$) potential (section  \ref{subsec:2}). There is no exchange potential involved in the positron scattering.

\textcolor{black}{The scattering parameters obtained in the body-fixed frame are then transformed to the space-fixed frame \cite{gianturco1986theory}.} Eventually, for non-polar molecules, the rotationally elastic (or rotationally resolved) DCS for the case where there is no transition between the initial and final rotational states ($n=0\rightarrow n'=0$) and the rotationally inelastic (or rotationally unresolved) DCS for transitions between different rotational states ($n\rightarrow n'$) is computed in the space-fixed frame as,

\begin{align}
 \frac{d\sigma}{d\Omega}(n\rightarrow n';\theta)= \sum_{L=0}^{L=LBIG}A_L(n\rightarrow n')P_L(cos\theta),
 \label{eqn6}
\end{align}
where $A_L$ is the expansion coefficient which depends upon the energy of the incoming projectile and is determined from the K-matrix elements, $LBIG$ is the maximum number of partial waves considered for the continuum positron, and $P_L(cos\theta)$ is the Legendre polynomial for a scattering angle $\theta$. The overall sum of the rotationally resolved and unresolved DCS gives the rotationally summed DCS.

When dealing with polar molecules, the long-range dipole interaction between the projectile and the target molecule leads to a slow convergence of the sum in Eq.\ref{eqn6}. To circumvent this challenge, the following closure formula is used,

 % \begin{widetext}
\begin{align}
 \frac{d\sigma}{d\Omega}(n\rightarrow n';\theta)= \frac{d\sigma^B}{d\Omega} +  \sum_{L=0}^{L=LBIG}(A_L - A_L^B)P_L(cos\theta),
 \label{eqn7}
\end{align}
 % \end{widetext}
where the ones with a superscript $B$ are calculated using the first Born approximation, resulting in a Born corrected DCS for polar molecules (Eq.\ref{eqn7}). The corresponding elastic ICS is computed using,

\begin{align}
 \sigma = \sigma_{rd}^B + \sigma_{cc} - \sigma_{fd}^B
 \label{eqn8}
\end{align}
Here, $\sigma_{rd}^B$ is the ICS calculated for a rotating dipole within the first Born approximation, $\sigma_{cc}$ is the result of close-coupling calculation under fixed-nuclei approximation, and $\sigma_{fd}^B$ is the ICS for a fixed dipole.

Similarly, the MTCS is calculated using,

\begin{align}
 \sigma^M= \sigma_{rd}^{M(B)} + \sigma_{cc}^M - \sigma_{fd}^{M(B)}
 \label{eqn9}
\end{align}

The general formula for calculating the K-matrix elements, DCS, ICS, and MTCS within the first Born approximation, as outlined by Sanna and Gianturco \cite{Polydcs}, is applicable only to molecules with higher symmetry (e.g., $\mathrm{C_{2v}}$) and a dipole moment aligned along the symmetry axis or non-polar molecules. For polar molecules with lower point group symmetries (e.g., $\mathrm{C_{1}}$ or $\mathrm{C_{s}}$), this formula requires modification, as suggested by Franz \textit{et al.} \cite{franz2014low} and implemented by us in our recent work \cite{snigdha_chemphychem}.

Please note that the ICS, DCS, and MTCS reported in this work are rotationally summed.

\subsection{Model potentials} \label{subsec:2}
As our motive is to highlight the significance of correlation in positron scattering dynamics and bound state formation with molecules, we have considered two models for the short-range correlation potential. The first is the one given by Perdew and Zunger \cite{Perdew}, 

\begin{align}
  V_{corr} (\textbf{r}) & = lnr_s(0.0311+0.00133r_s) - 0.0084r_s - 0.0584 \qquad \rm{for} \; r_s < 1.0 ,\nonumber \\
                     & = \frac{\gamma(1+\frac{7}{6}\beta_1r_s^{1/2}+\frac{4}{3}\beta_2r_s)}{(1+\beta_1r_s^{1/2}+\beta_2r_s)^2} \qquad\rm{for} \;r_s \geq 1.0,
\end{align}
where $\gamma$ = -0.1423, $\beta_1$ = 1.0529, $\beta_2$ = 0.334, and $r_s$ = $[\frac{3}{4\pi\rho(\textbf{r})}]^{1/3}$. This potential will be represented as $\rm{V_{cp}}$ in this work.

The second one is a recent model developed by Swann and Gribakin \cite{Swann2018} as an extension of the positron-atom correlation given by Mitroy and Ivanov \cite{Mitroy_Ivanov}. It is a free parameter model that takes into account the other correlations, like virtual positronium formation, and is given as,         

\begin{align}
    V_{corr} (\textbf{r})= - \sum_{A=1}^{N_a} \frac{\alpha_A}{2|\textbf{r}-\textbf{r}_A|^4} \left[1-exp\left(-\frac{|\textbf{r}-\textbf{r}_A|^6}{\rho_A^6}\right)\right].
\end{align}

Here, $\alpha_A$ is the hybrid polarizability of the individual atoms, $A$, of the molecule as given by Miller \cite{Miller}. It accounts for the chemical environment of the atoms in the molecule. $\rho_A$, the cut-off radius, is a free parameter whose value is obtained for each atom by a fitting procedure elaborated in section \ref{subsec:level4A}. $\textbf{r}$ denotes the position vector of the positron and $\textbf{r}_A$ is the position vector of the atom $A$. Owing to its success in reproducing the positron bound states with atoms and molecules \cite{swann2020model,swann2021effect, frighetto2024low}, we have integrated this model potential to the codes used to implement the SCE method and will be represented as $\rm{V_{sg}}$.

Further, for the long-range polarization potential, the given form, which depends upon the \textcolor{black}{dipolar} polarizability $\alpha$ of the target molecule \cite{SCElib}, is used,

\begin{align}
  V_{pol} (\textbf{r})  = - \frac{\alpha}{2r^4},
\end{align}
where $\alpha$ = $\sum_A \alpha_A$. Anisotropy of the atomic polarizability and the contribution to the polarization potential from the higher-order terms are neglected.   

\subsection{Scattering length and virtual/bound state energies}

Positron virtual or bound state formation can be estimated by calculating the scattering length (A), given as \cite{Morrison},

\begin{align}
  A =-\lim_{k\rightarrow0}\frac{tan[\delta_0(k)]}{k},
\end{align}
where $\delta_0$ is the s-wave eigenphase for positron momentum $k$. For small $k$ values, utilizing the effective-range theory expansion \cite{spruch1960modification,swann2020model}, we can write,

\begin{align}
  k\cot\delta_0 =-\frac{1}{A} + \frac{\pi\alpha}{3A^2}k+O(k^2lnCk),
  \label{eq14}
\end{align}
$\alpha$ being the polarizability of the molecule. The fit for Eq. \ref{eq14} \cite{swann2020model}, used in this work to calculate the scattering length $A$, is,

\begin{align}
  k\cot\delta_0 =-\frac{1}{A} + Ck,
  \label{eq15}
\end{align}
where $C$ is the fitting parameter. The lowest two values of positron momentum, closer to zero, and the corresponding $\delta_0$ values were used to estimate $A$ and $C$, as suggested by Morrison \cite{Morrison}. Positive $A$ denotes a bound state, whereas negative $A$ corresponds to a virtual state. 

Ultimately, the virtual or bound state energy ($\epsilon$) can be calculated using \cite{frighetto2023low,frighetto2024low},

\begin{align}
  \epsilon \approx \frac{1}{2A^2}
  \label{eq15}
\end{align}

\section{\label{sec:level3}Computational details\protect}

% \begin{table*}
% \centering
% \caption{\label{tab:table1}Comparison of the SCE computed and the theoretical/experimental dipole moment of targets (Units in Debye). }
% % \begin{ruledtabular}
% \begin{tabular}{ccp{3cm}p{3cm}c}\hline
%  % &\multicolumn{2}{c}{$D_{4h}^1$}&\multicolumn{2}{c}{$D_{4h}^5$}\\
% Target&Point group&Computed number of electrons (SCE)&Computed dipole moment (SCE) 
% &Compared dipole moment\\ \hline
% Benzene\\ $(C_6H_6)$  & $D_{6h}$ & 42.097 & 0.0 & 0.0 \cite{g16}, 0.0 \cite{NIST} \\
% Furan\\ $(C_4H_4O)$  & $C_{2v}$ & 36.005 & 0.74981 & 0.7714 \cite{g16}, 0.660 \cite{NIST} \\
%  2H-pyran\\ $(C_5H_6O)$&$C_{s}$&44.005 &0.90946 & 0.89810 \cite{g16}, 0.94 \cite{silva2024elastic} \\
% 4H-pyran\\ $(C_5H_6O)$&$C_{2v}$ & 44.011 & 0.92017 & 0.91500 \cite{g16}, 1.14 \cite{silva2024elastic}\\ 
% Pyrrole\\ $(C_4H_4NH)$  & $C_{2v}$ & 36.003 & 1.89081 & 1.8905 \cite{g16}, 1.84 \cite{NIST} \\\hline
% \end{tabular}
% % \end{ruledtabular}
% \end{table*}

The ground state geometries of the target molecules were optimized using Density Functional Theory (DFT) within the Gaussian 16 software package \cite{g16}. Molecular symmetries were maintained at their natural point groups: $\mathrm{D_{2h}}$ for ethylene, $\mathrm{C_{2v}}$ for water, and $\mathrm{C_{s}}$ for formic acid. In the case of linear molecules, oxygen, hydrogen, and acetylene, the abelian point group $\mathrm{D_{2h}}$ was considered. The electronic structure and ground state molecular orbitals were determined using the self-consistent field (SCF) method within the DFT framework. The PBEPBE functional in conjunction with the aug-cc-pVDZ basis set was utilized for hydrogen molecule, water, and formic acid. For oxygen molecule, the B3LYP functional was employed. To facilitate direct comparison with existing literature, the wB97XD functional with the 6-311++G(3d,3p) basis set was used for ethylene and acetylene.

\begin{table*}[h]
\centering

% \begin{ruledtabular}
\begin{tabular}{|c|c|c|c|}
 % &\multicolumn{2}{c}{$D_{4h}^1$}&\multicolumn{2}{c}{$D_{4h}^5$}\\
\hline
Target & Exp. dipole moment (Debye) \cite{NIST} &  Polarizability (a.u.) \cite{Miller}  \\ \hline

Hydrogen molecule $(H_2)$ & 0.0 & 5.224  \\ \hline

Ethylene $(C_2H_4)$ & 0.0 & 28.696 \\ \hline

Acetylene $(C_2H_2)$ & 0.0 & 22.54 \\ \hline

Oxygen molecule $(O_2)$ & 0.0 & 7.688 \\ \hline
Water $(H_2O)$ & 1.86 & 9.527 \\ \hline
Formic acid ($HCOOH$) & 1.410 & 22.495 \\ \hline
\end{tabular}
\label{table1}
\caption{\label{tab:table1a}Molecular properties of the targets used in the present calculations.}
% \end{ruledtabular}
\end{table*}

The bound wavefunctions, electron densities, and potentials were then expanded about a single center in the basis of angular functions using \textcolor{black}{SCELib 4.0} \cite{SCElib}, a set of codes developed by Sanna \textit{et al.} To ensure satisfactory convergence of the molecular properties, the maximum number of partial waves, $l_{max}$, in Eq.\ref{eqn1} was set to 50 for the targets, normalizing the molecular orbitals to approximately unity. Radial and angular grids chosen for integration provided accurate calculations of the number of electrons and gave good values for the dipole moments (if any), indicating satisfactory modeling of the target molecules. The sum of the hybrid polarizabilities of individual atoms, as taken from \cite{Miller}, gave the total isotropic polarizability of the molecule. The reason for including only isotropic polarizability was to align with the calculations done in the literature and provide a true comparison with the data.  Polarizability is critical for accurately modeling the polarization potential in asymptotic regions; hence, the anisotropy involved will be included in our future work. The dipole moment and polarizability values used in the calculations are summarized in Table \ref{tab:table1a}.

The next step in our calculation involves solving the coupled radial equations for scattering (Eq.\ref{eqn4}) using the VOLSCAT 2.0 code \cite{Volscat}. This yields the k-matrices, which are used to calculate the ICS. To ensure convergence of the result, we considered a sufficient number of partial waves in the scattering calculation. However, the determined ICS is initially in the body-fixed frame. To obtain the DCS and ICS in the space-fixed frame, we employed the POLYDCS code \cite{Polydcs}. For the frame transformation, we utilized the ASYMTOP code \cite{JAIN1983301} to generate the molecules' rotational energy levels and eigenfunctions, except for symmetric molecules for which the rotational levels are trivial. Specifically, we considered the first six rotational transitions ( $\rm{0\rightarrow0, 0\rightarrow1, 0\rightarrow2, 0\rightarrow3, 0\rightarrow4, and \: 0\rightarrow5}$) for the molecules.
 Furthermore, the original POLYDCS code suite \cite{Polydcs} considers molecules with either an axis of symmetry aligned with the dipole moment or a zero dipole moment. Since formic acid belongs to the lower point group $\mathrm{C_{s}}$, we employed our modified POLYDCS code for it.

\section{\label{sec:level4}Results and discussion\protect}

\subsection{\label{subsec:level4A}Optimization of cut-off radius ($\rho_A$)}
\begin{figure*}
    \centering
    \subfigure[Hydrogen molecule]{%
        \includegraphics[width=8cm, height=10cm, keepaspectratio]{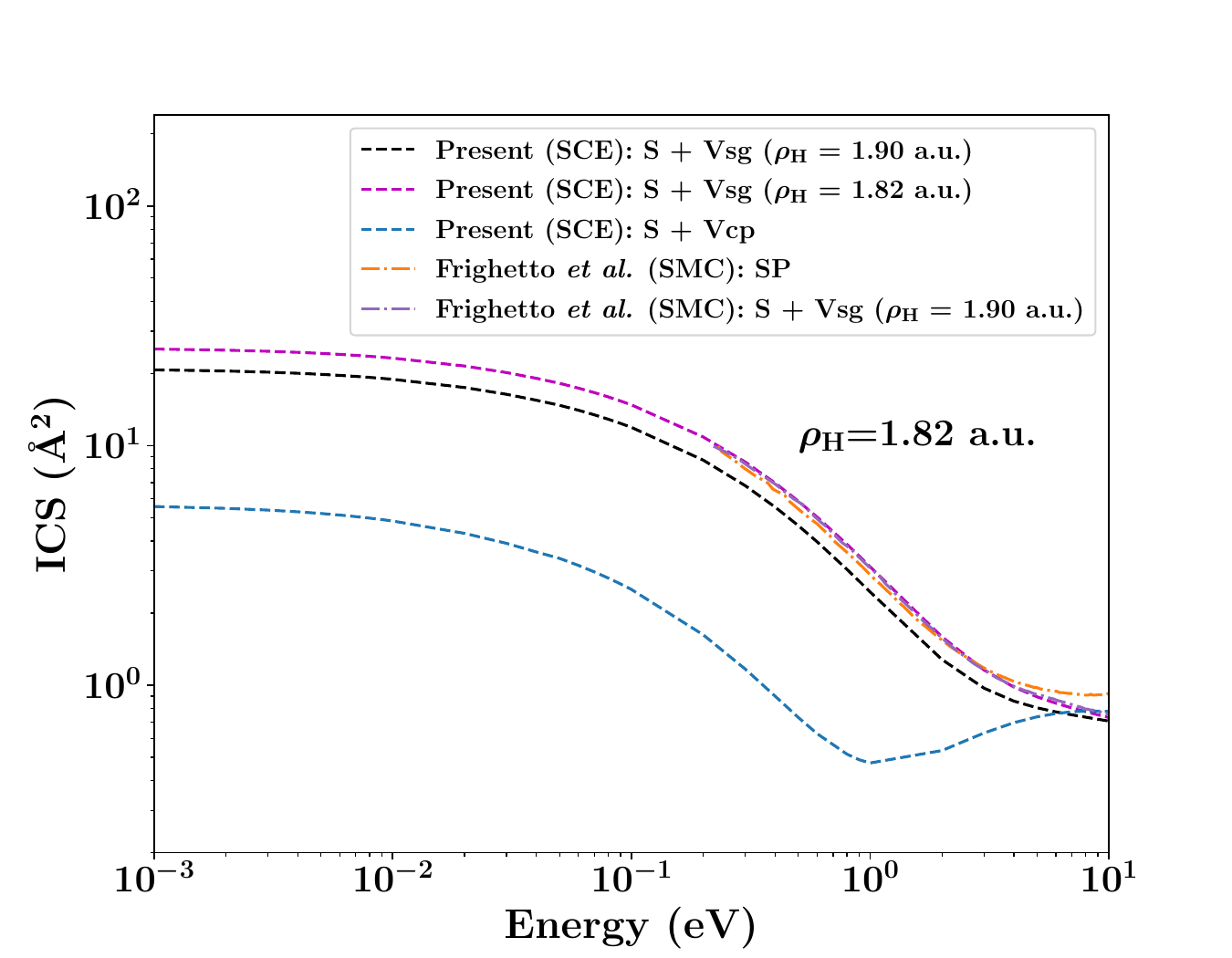}
        \label{fig1:subA}
    }
    \subfigure[Ethylene (\rm{$sp^2$} carbon)]{%
        \includegraphics[width=8cm, height=10cm, keepaspectratio]{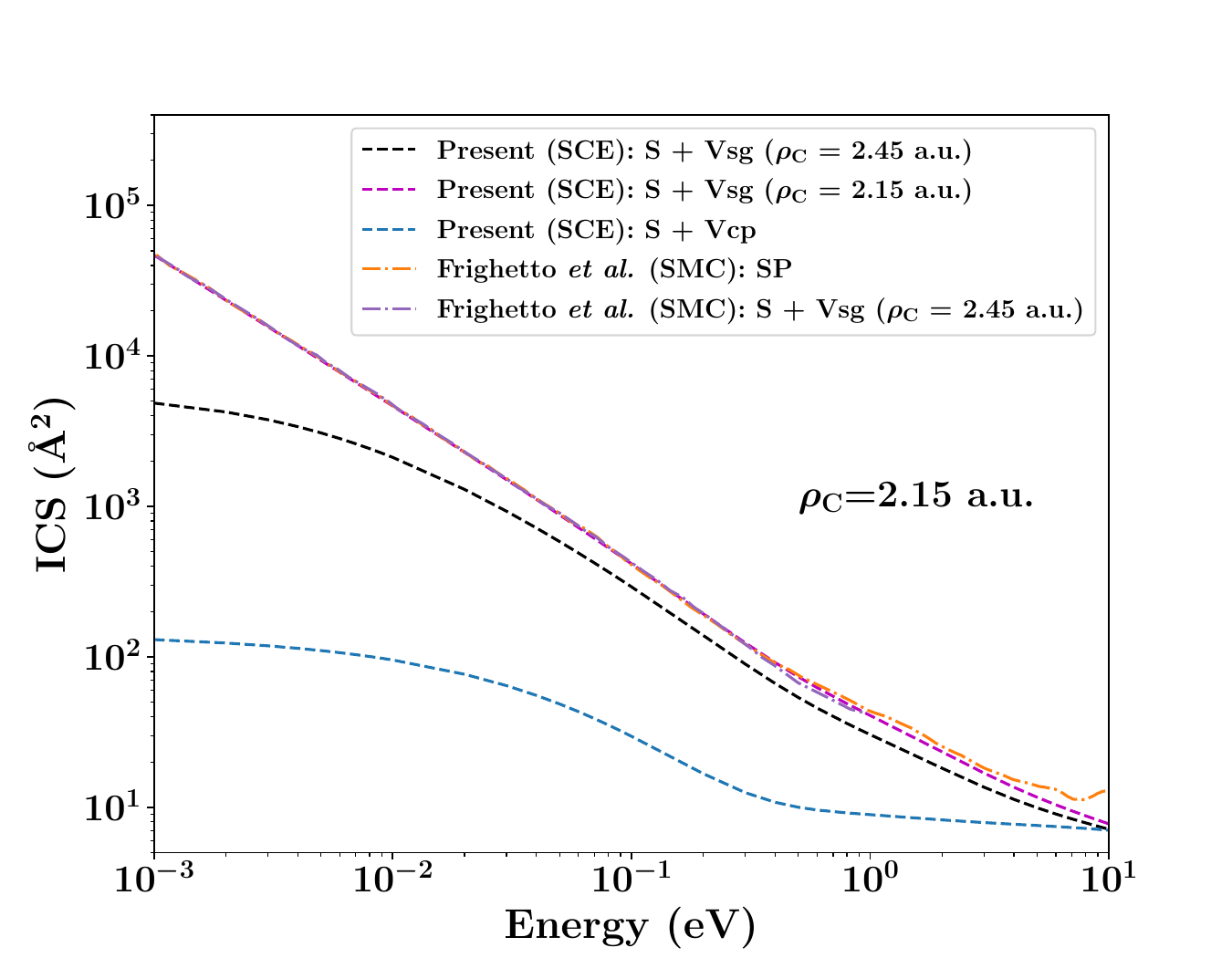}
        \label{fig1:subB}
    }
    \subfigure[Acetylene (\rm{$sp$} carbon)]{%
        \includegraphics[width=8cm, height=10cm, keepaspectratio]{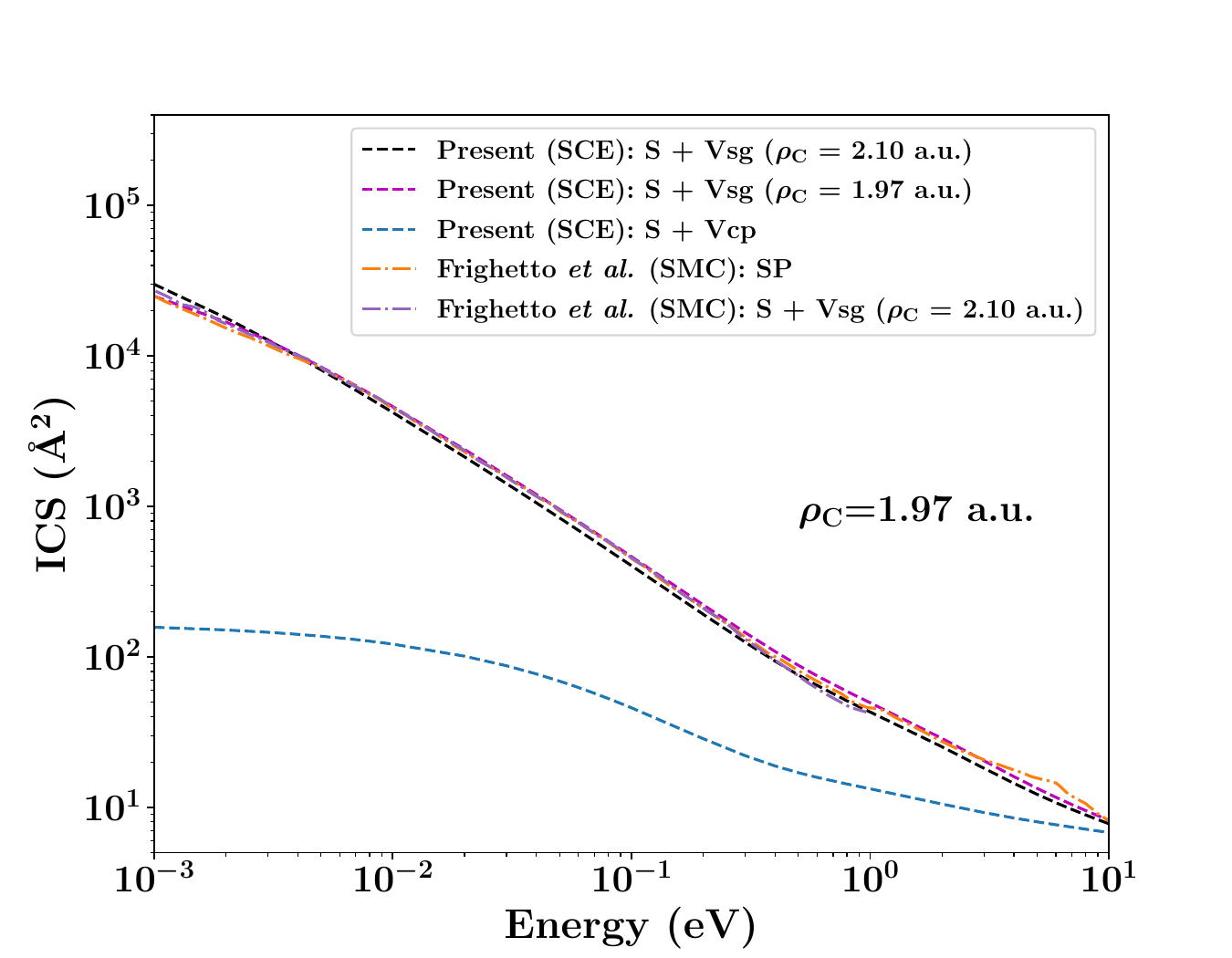}
        \label{fig1:subC}
    }
    \subfigure[Oxygen molecule (\rm{$sp^2$} oxygen)]{%
        \includegraphics[width=8cm, height=10cm, keepaspectratio]{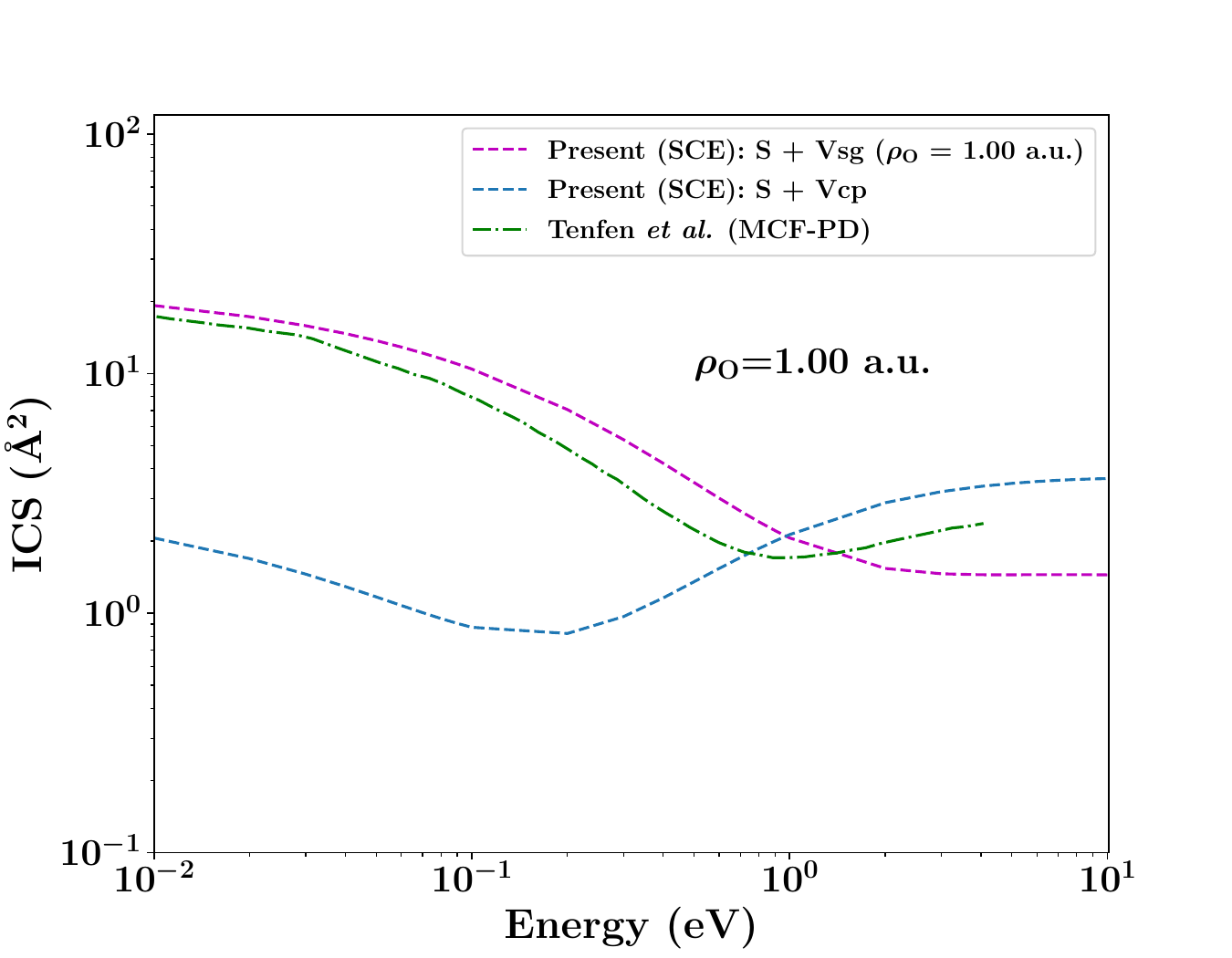}
        \label{fig1:subD}
    }
    \subfigure[Water (\rm{$sp^3$} oxygen)]{%
        \includegraphics[width=8cm, height=10cm, keepaspectratio]{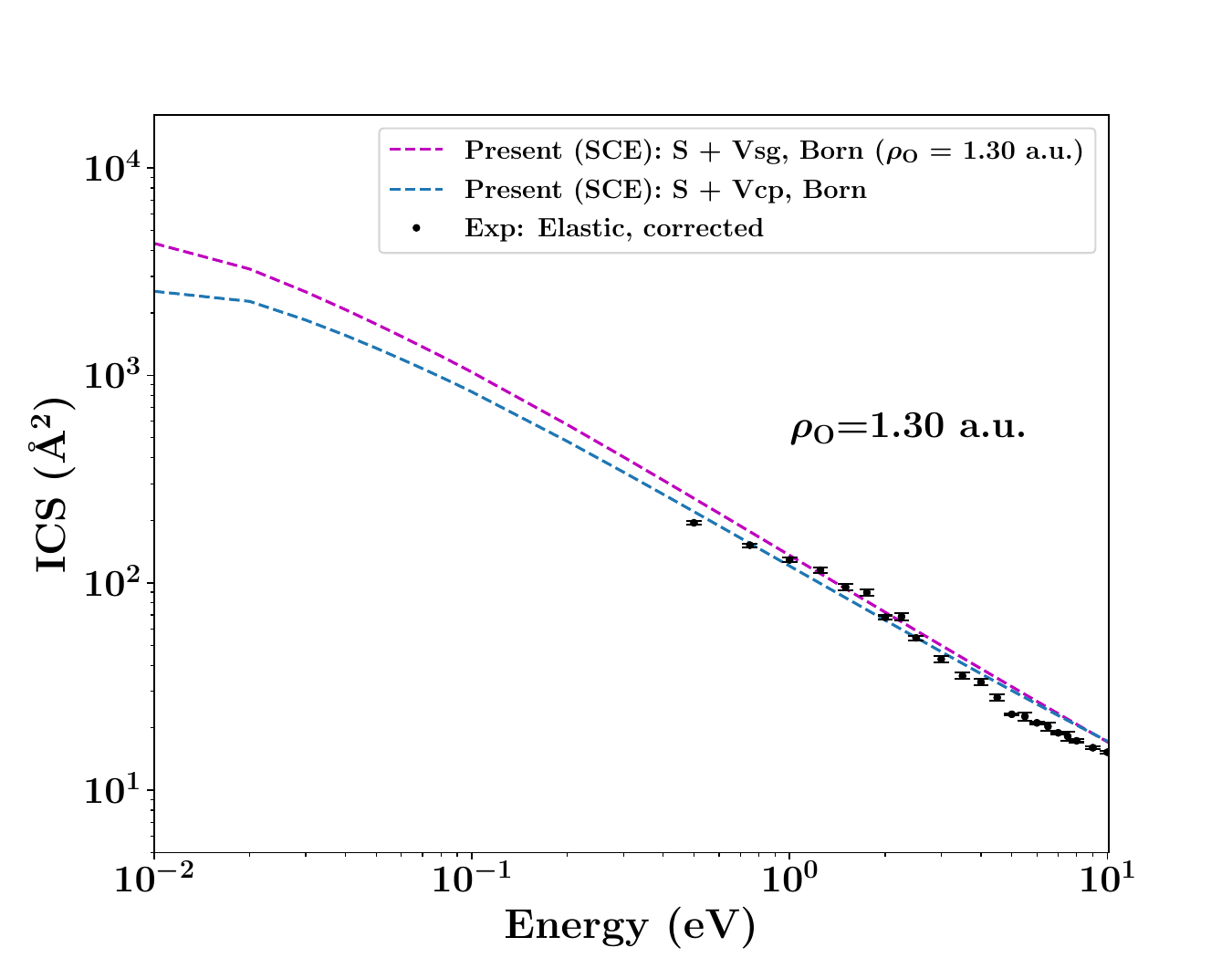}
        \label{fig1:subE}
    }
    % \captionsetup{width=1.0\textwidth}
     
\end{figure*}

The optimized value of the cut-off radius, $\rho_A$, for different atoms was obtained by directly fitting the calculated ICS for smaller molecules (containing the desired atoms) with the best available experimental or \textit{ab initio} data. We started with hydrogen atom, for which $\rho_H$ was estimated by fitting the ICS calculated for hydrogen molecule, using $\rm{V_{sg}}$ correlation, with the data reported by Frighetto \textit{et al.} \cite{Frighetto2023Imp} as shown in Fig. \ref{fig1:subA}. They employed the best of \textit{ab initio} SMC calculations (SP approximation) as well as the model polarization (S+$\rm{V_{sg}}$) approach. In our case, the $\rho_H$ value of 1.82 a.u. gave an excellent agreement with the ICS of Frighetto, both SP and S+$\rm{V_{sg}}$. The $\rho_H$ value is a bit lower than that used by Frighetto \textit{et al.} (i.e., 1.90 a.u.). This variation is expected, as in the SCE approach, the model potential is expanded about a single center, which is not the case in the SMC method. To provide a clear picture, we have also illustrated the ICS obtained by considering $\rho_H$ as 1.90 a.u., which shows deviation from the compared data. While the ICS obtained by considering the $\rm{V_{sg}}$ potential is at par with the best of \textit{ab initio} calculations, the one obtained by using the $\rm{V_{cp}}$ potential is a lot off. 

\begin{table*}[h]
\centering

% \begin{ruledtabular}
\begin{tabular}{|>{\centering\arraybackslash}p{4cm}|>{\centering\arraybackslash}p{4cm}|c|c|}
 % &\multicolumn{2}{c}{$D_{4h}^1$}&\multicolumn{2}{c}{$D_{4h}^5$}\\
\hline
Atom & $\rho_A$ in a.u.  \\ \hline

H & 1.82  \\ \hline

C $(sp^2)$ & 2.15 \\ \hline

C $(sp)$ & 1.97 \\ \hline

O $(sp^2)$ & 1.00 \\ \hline
O $(sp^3)$ & 1.30 \\ \hline

\end{tabular}
\label{table1}
\caption{\label{tab:table2a}Cut-off radius ($\rho_A$) determined for the constituent atoms of the targets.}
% \end{ruledtabular}
\end{table*}

For $\rm{sp^2}$ hybridized carbon atom, we calculated the ICS for ethylene. We fitted the data with the result reported by Frighetto \textit{et al.} \cite{frighetto2024low,frighetto2023low}, calculated both with SP approximation and S+$\rm{V_{sg}}$ SMC approach, as shown in Fig. \ref{fig1:subB}. This gave us a $\rho_C$ value of 2.15 a.u. Again, the ICS calculated by us using the $\rho_C$ value (2.45 a.u.) reported in Frighetto \textit{et al.}'s work deviates from their data. Furthermore, the ICS calculated using the $\rm{V_{cp}}$ potential again shows a marked difference, specifically towards the lower energies. The fitting of ICS for acetylene, which has two $\rm{sp}$ hybridized carbon atoms, with the data reported by Frighetto \textit{et al.} \cite{frighetto2024low,frighetto2023low}\textemdash both SP and S+$\rm{V_{sg}}$\textemdash resulted in the $\rho_C$ value of 1.97 a.u., again lower than that used by Frighetto (2.10 a.u.). Similar to the case of hydrogen molecule and ethylene, ICS obtained for acetylene using the $\rm{V_{cp}}$ potential shows a substantial difference, as illustrated in Fig. \ref{fig1:subC}.  

Moving further, for $\rm{sp^2}$ hybridized oxygen atom, we considered oxygen molecule and fitted the evaluated ICS with the data reported by Tenfen \textit{et al.} \cite{TENFEN2022} employing the method of continued fractions (MCF) using dipole polarizability (PD). The best fitting we could obtain was for a $\rho_O$ value of 1.0 a.u. We can see in Fig. \ref{fig1:subD}, the $\rm{V_{sg}}$ ICS compares well with the literature data for positron energies roughly below 2 eV, above which, the $\rm{V_{cp}}$ ICS gives a better agreement. Fitting the calculated ICS for water with the experimental elastic cross-section, obtained by subtracting the sum of excitation and ionization cross-sections of Tattersall \textit{et al.} \cite{Tattersall} from the grand total cross-section reported by Makochekanwa \textit{et al.} \cite{Makochekanwa_2009}, gave us a $\rho_O$ value of 1.30 a.u. for $\rm{sp^3}$ hybridized oxygen atom. Note that the experimental cross-section of Makochekanwa \textit{et al.} is corrected for forward scattering effects, providing better data for comparison. 

In general, we can comment that the $\rm{V_{cp}}$ correlation results in the ICS data that significantly deviates---both quantitatively and qualitatively---from the data obtained using the $\rm{V_{sg}}$ correlation, towards lower energies for non-polar molecules. For polar molecules, like water in this case, the deviation is comparatively smaller. This is due to the significant contribution of the dipole moment in the low-energy scattering dynamics. For a detailed discussion, refer to the section \ref{subsec:level4B}. The optimized cut-off radius for the atoms of the different molecules is shown in Table \ref{tab:table2a}.  These cut-off radii were ultimately used to perform scattering calculations for formic acid, which has two hydrogen atoms, one $\rm{sp^2}$ hybridized oxygen atom, one $\rm{sp^3}$ hybridized oxygen atom, and a $\rm{sp^2}$ hybridized carbon atom.

\subsection{\label{subsec:level4B}Elastic cross-sections}
\subsubsection{Oxygen molecule}

\begin{figure}[h]
    \centering
    \subfigure[]{%
        \includegraphics[width=8cm, height=10cm, keepaspectratio]{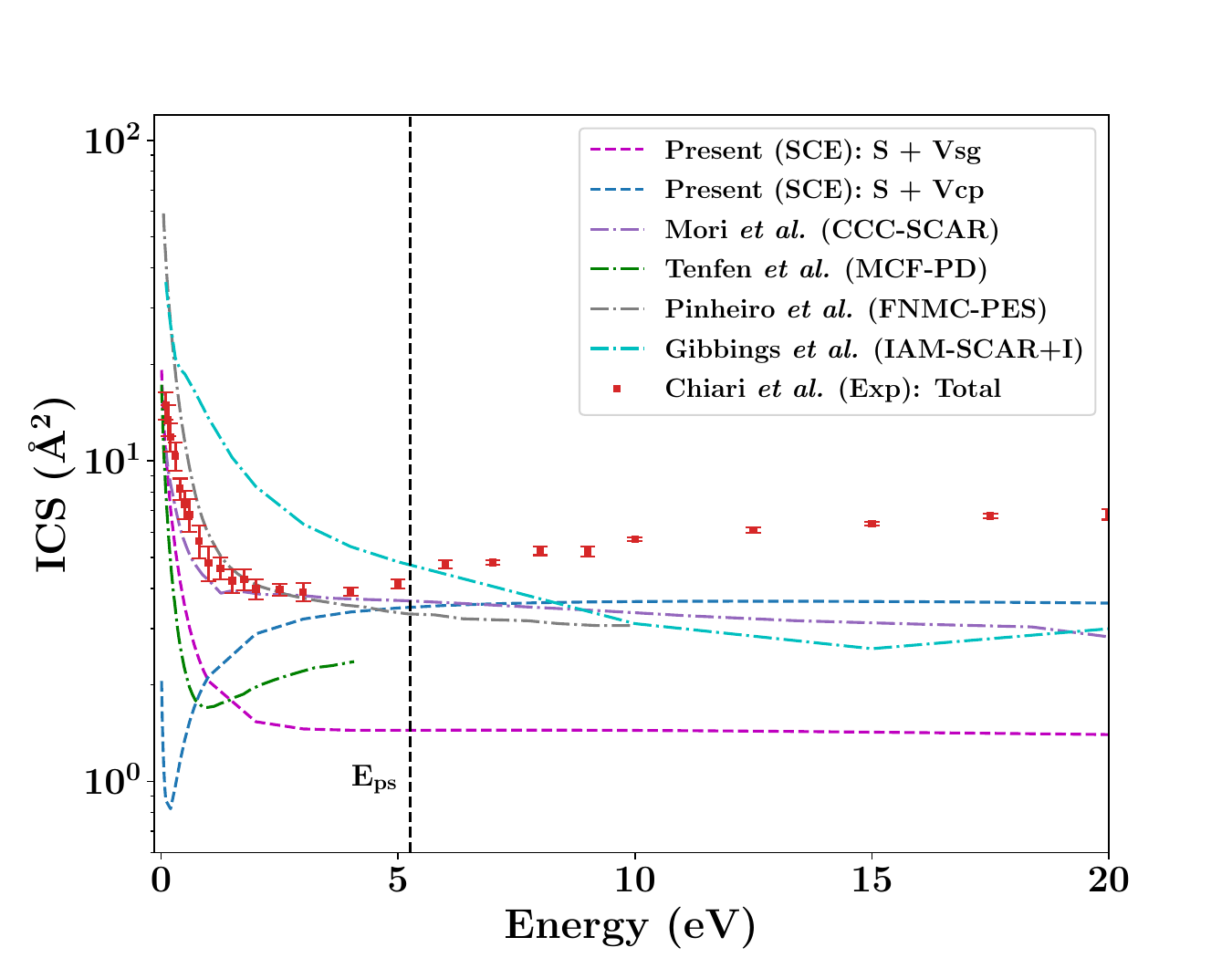}
        \label{fig2:subA}
    }
    \subfigure[]{%
        \includegraphics[width=8cm, height=10cm, keepaspectratio]{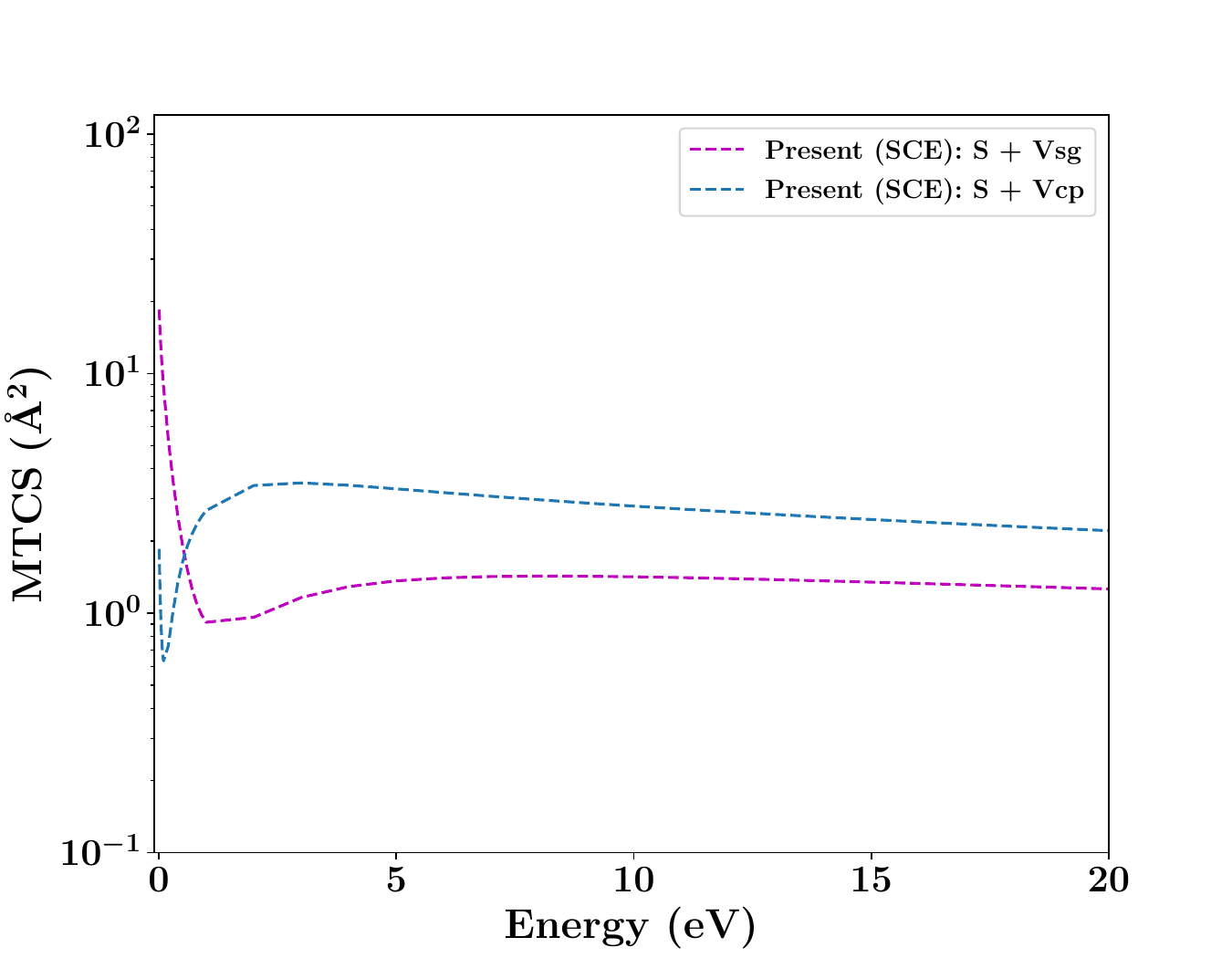}
        \label{fig2:subB}
    }
    \caption{(a) Integral cross-section and (b) momentum transfer cross-section for positron scattering from oxygen molecule. Magenta dashed line: present calculation using $V_{sg}$ correlation. Blue dashed line: present calculation using $V_{cp}$ correlation. Purple dash-dot line: CCC-SCAR data of Mori \textit{et al.} \cite{mori2024calculations}. Green dash-dot line: MCF-PD data of Tenfen \textit{et al.} \cite{TENFEN2022}. Grey dash-dot line: Pinheiro \textit{et al.} \cite{pinheiro2023elastic} data. Cyan dash-dot line: Gibbings \textit{et al.} \cite{Gibbings_2019} IAM-SCAR+I data. Red squares: experimental total cross-section of Chiari \textit{et al.} \cite{Chiari_2012}.  }
    \label{fig:2}
\end{figure}

Fig. \ref{fig2:subA} represents the ICS of positron scattering from oxygen molecule. The incident energy of the positron is considered up to 20 eV. Results involving both $\rm{V_{sg}}$ and $\rm{V_{cp}}$ potentials are shown in the figure. While the $\rm{V_{cp}}$ ICS deviates significantly from the general trend of the cross-section towards lower energies, the $\rm{V_{sg}}$ ICS compares satisfactorily with other theoretical and experimental results. The magnitude of the latter is consistently below the compared results, approximately above 2 eV, giving a qualitative agreement rather than a quantitative one. One of the reasons is the $\rho_O$ value, which is determined by fitting very low energy to 10 eV cross-section data. As higher partial waves also contribute when energy increases, the cut-off radius should vary for different $l$ values, as suggested by Mitroy and Ivanov \cite{Mitroy_Ivanov}. Below 2 eV, our $\rm{V_{sg}}$ ICS agrees well with the data of Tenfen \textit{et al.} \cite{TENFEN2022}; also discussed in the subsection \ref{subsec:level4A}. They employed the method of continued fractions (MCF) using dipole polarizability (PD) as well as higher-order polarizability (PG) terms for the polarization potential. We have compared our result with the PD ICS since we have also considered only the dipole polarizability for the asymptotic polarization potential. However, the polarizability value used in our case is 7.688 a.u. (sum of hybrid polarizabilities), which is lower than that considered by Tenfen \textit{et al.} (10.46 a.u.). This is evident in the cross-section values above 2 eV, where Tenfen's data rises above our ICS. This work aims to show the effect of short-range correlation in the simplest form. The higher-order polarizabilities, anisotropy, as well as different cut-off radii for different partial waves will be considered in our future work.    

Below roughly 1 eV, our $\rm{V_{sg}}$ ICS compares well with the recent data of Mori \textit{et al.} \cite{mori2024calculations} as well as the total cross-section (TCS) of Chiari \textit{et al.}\cite{Chiari_2012}. The TCS data of Chiari \textit{et al.} are uncorrected for the forward scattering effects, which can be neglected as the oxygen molecule is nonpolar \cite{mori2024calculations}. Mori \textit{et al.} employed the CCC-SCAR technique in which the convergent close-coupling (CCC) results for atomic targets were used to perform calculations under the framework of the independent atom model with screening-corrected additivity rule (IAM-SCAR). This implementation is expected to give a better result than using the atomic cross-sections from model potential techniques. Qualitatively, Mori's data are in good agreement with our $\rm{V_{sg}}$ ICS, where, the overall higher magnitude of the former can be attributed to the implementation of IAM, which generally results in higher cross-section values. The IAM-SCAR+I (Independent atom model with screening corrected additivity rule plus interference terms) data of Gibbings \textit{et al.} \cite{Gibbings_2019} are consistently higher than ours, all throughout the energy range. This is again expected as the IAM-SCAR method is known to result in higher values of cross-sections. The ICS data reported by Pinheiro \textit{et al.} \cite{pinheiro2023elastic}, implementing the FNMC (finite nuclear mass correction) technique to construct the potential energy surface (PES) together with the close-coupling formalism, qualitatively agree with our result, but are again higher in magnitude. Please note that our calculation utilizes only the dipolar polarizability for the asymptotic polarization. The hybrid polarizability of the atoms also neglects the anisotropy. These will affect the magnitude of the cross-section and will be considered in our future work. Also, oxygen being a non-polar molecule with low polarizability, totally reflects the dependence of the ICS on the correlation-polarization potential towards the lower energies. This is evident from the large deviation of our $\rm{V_{cp}}$ ICS from the $\rm{V_{sg}}$ ICS, as well as other compared data, in both magnitude and form. The dip observed in the $\rm{V_{cp}}$ cross-section, particularly seen in the case of non-polar molecules, reflects the unsatisfactory modeling of the correlation-polarization potential \cite{Tenfen_2019}, which vanishes in the $\rm{V_{sg}}$ and other compared cross-sections.  Above the positronium formation threshold of 5.263 eV \cite{Chiari_2012}, the $\rm{V_{cp}}$ ICS gives a better comparison with literature as towards higher energies, the effect of correlation-polarization diminishes.

MTCS for the oxygen molecule is presented in Fig. \ref{fig2:subB}. Here again, the $\rm{V_{cp}}$ data deviates from the $\rm{V_{sg}}$ data. The variation is both quantitative and qualitative below approximately 5 eV and only quantitative above that. There are no MTCS data in the literature for molecular oxygen, to the best of our knowledge. 

\begin{figure}
    \centering
    \subfigure[]{%
        \includegraphics[width=8cm, height=10cm, keepaspectratio]{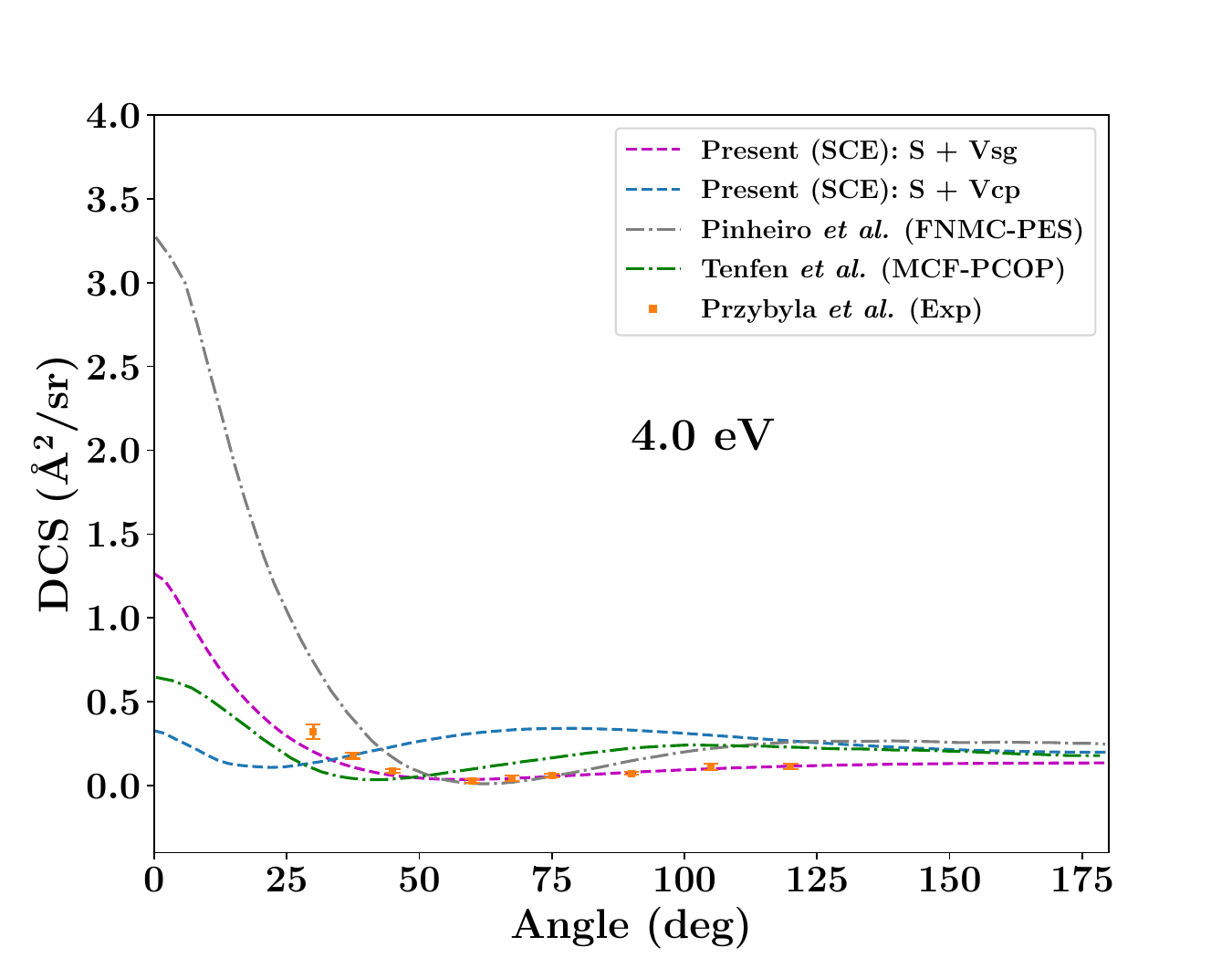}
        \label{fig3:subA}
    }
    \subfigure[]{%
        \includegraphics[width=8cm, height=10cm, keepaspectratio]{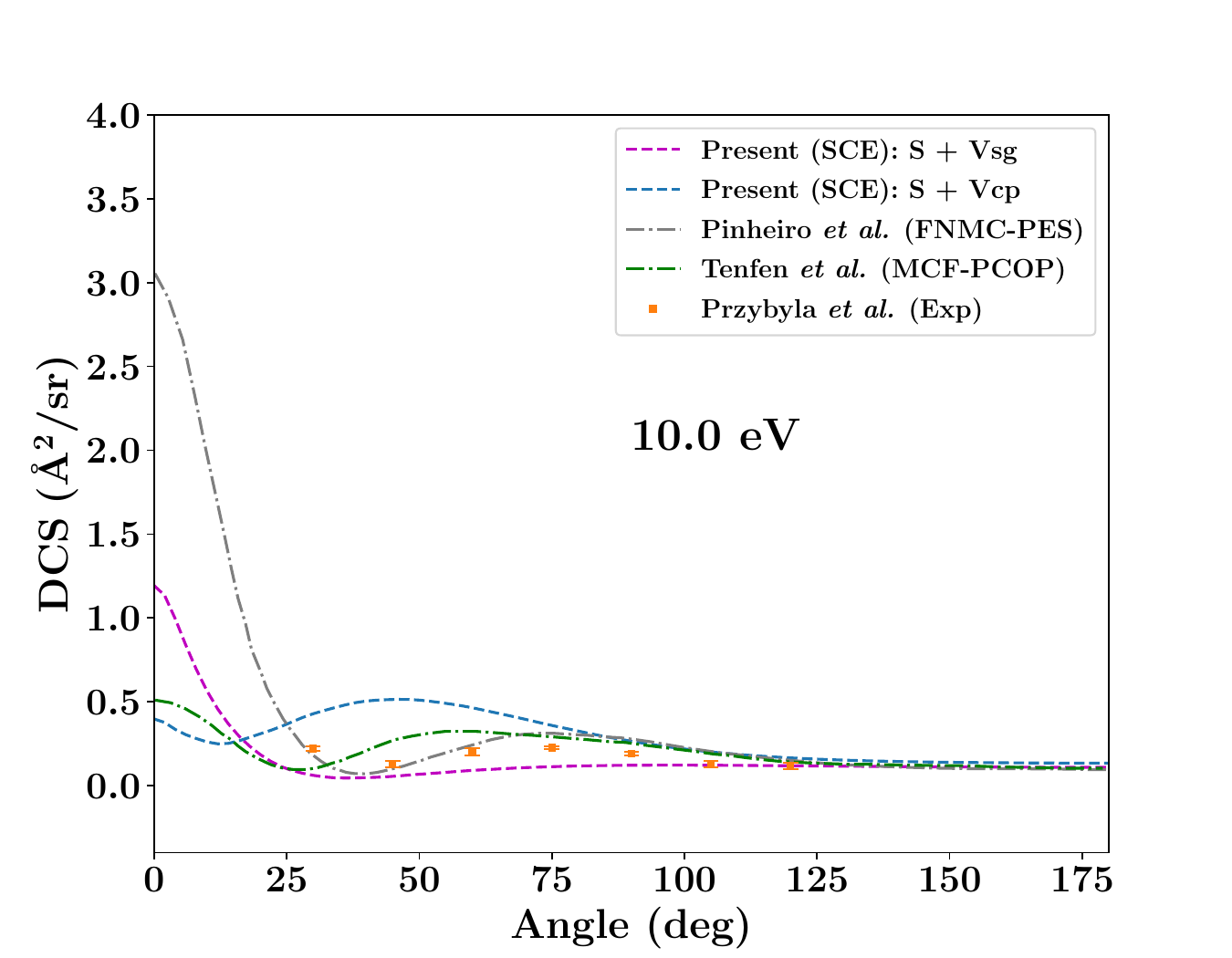}
        \label{fig3:subB}
    }
    \subfigure[]{%
        \includegraphics[width=8cm, height=10cm, keepaspectratio]{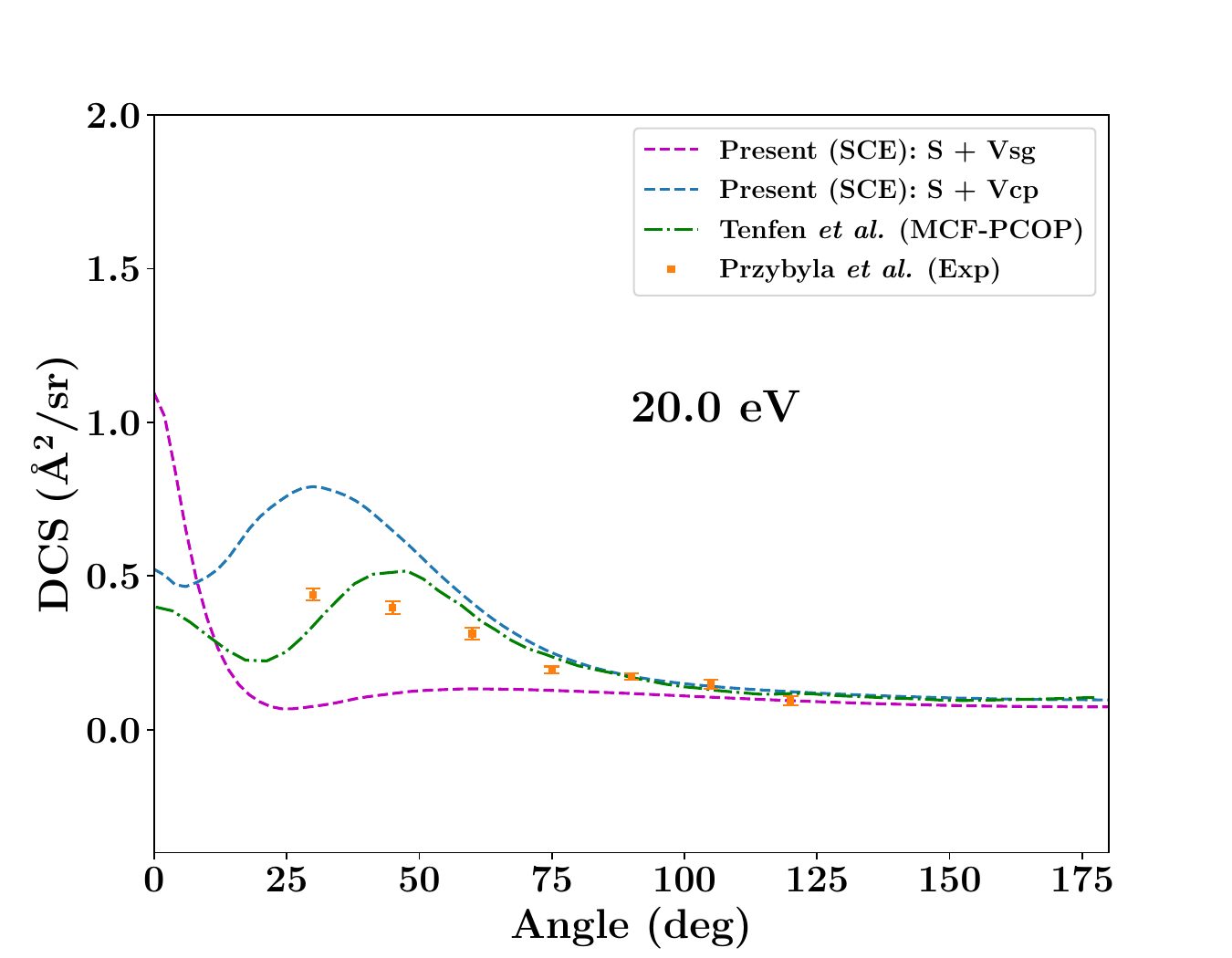}
        \label{fig3:subC}
    }
    \caption{Differential cross-section for positron scattering from oxygen molecule. Magenta dashed line: present calculation using $V_{sg}$ correlation. Blue dashed line: present calculation using $V_{cp}$ correlation. Grey dash-dot line: Pinheiro \textit{et al.} \cite{pinheiro2023elastic} data. Green dash-dot line: MCF-PCOP data of Tenfen \textit{et al.} \cite{Tenfen_2019}. Orange squares: experimental measurement of Przybyla \textit{et al.} \cite{Przybyla}.}
    \label{fig:3}
\end{figure}

The most stringent test of our correlation implementation will be provided by the DCS results, as shown in Fig. \ref{fig:3}. The only available experimental DCS for molecular oxygen is that of Przybyla \textit{et al.} \cite{Przybyla}. They have reported the relative DCS in arbitrary units. In order to provide a reasonable comparison, we normalized the experimental DCS at $\rm{120^0}$ with our $\rm{V_{sg}}$ DCS value. The $\rm{V_{sg}}$ DCS at 4 eV energy is in good agreement with the normalized experimental DCS, illustrated in Fig. \ref{fig3:subA}. It also compares fairly with the DCS of Tenfen \textit{et al.} \cite{Tenfen_2019}. Our data deviates from the result of Pinheiro \textit{et al.} \cite{pinheiro2023elastic} at forward angles, where the correlation-polarization plays a major part. The $\rm{V_{cp}}$ DCS features a different form below $\rm{125^0}$, above which it merges with the compared data. At 10 eV (refer to Fig. \ref{fig3:subB}), our $\rm{V_{sg}}$ DCS again shows a decent comparison with the experimental data of Przybyla, highlighting a similar form to that of Tenfen's and Pinheiro's data below $50^0$. The $\rm{V_{cp}}$ DCS once again is out of form towards lower angles, merging with the other data at higher angles. As we move to higher energies, like 20 eV in Fig. \ref{fig3:subC}, the $\rm{V_{sg}}$ correlation is unable to bring out the features of DCS as seen in the case of Tenfen as well as our $\rm{V_{cp}}$ data. The experimental data of Przybyla also deviates from the $\rm{V_{sg}}$ DCS at forward angles. This might be due to the increasing contribution of the higher partial waves, which can result in a different cut-off radius value for the $\rm{V_{sg}}$ correlation potential.

\subsubsection{Water}
\begin{figure*}
    \centering
    \subfigure[]{%
        \includegraphics[width=8cm, height=10cm, keepaspectratio]{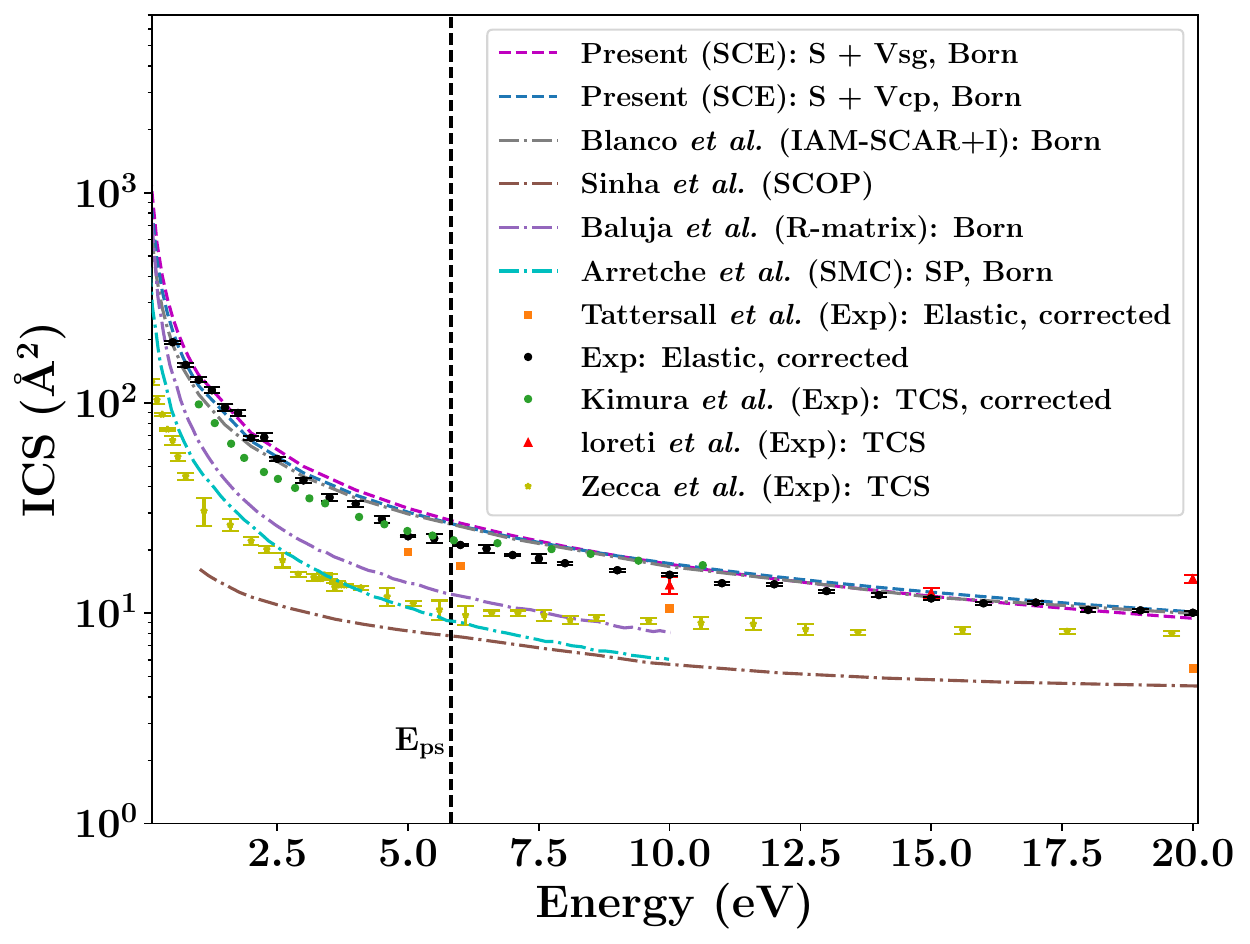}
        \label{fig4:subA}
    }
    \hfill
    \subfigure[]{%
        \includegraphics[width=8cm, height=10cm, keepaspectratio]{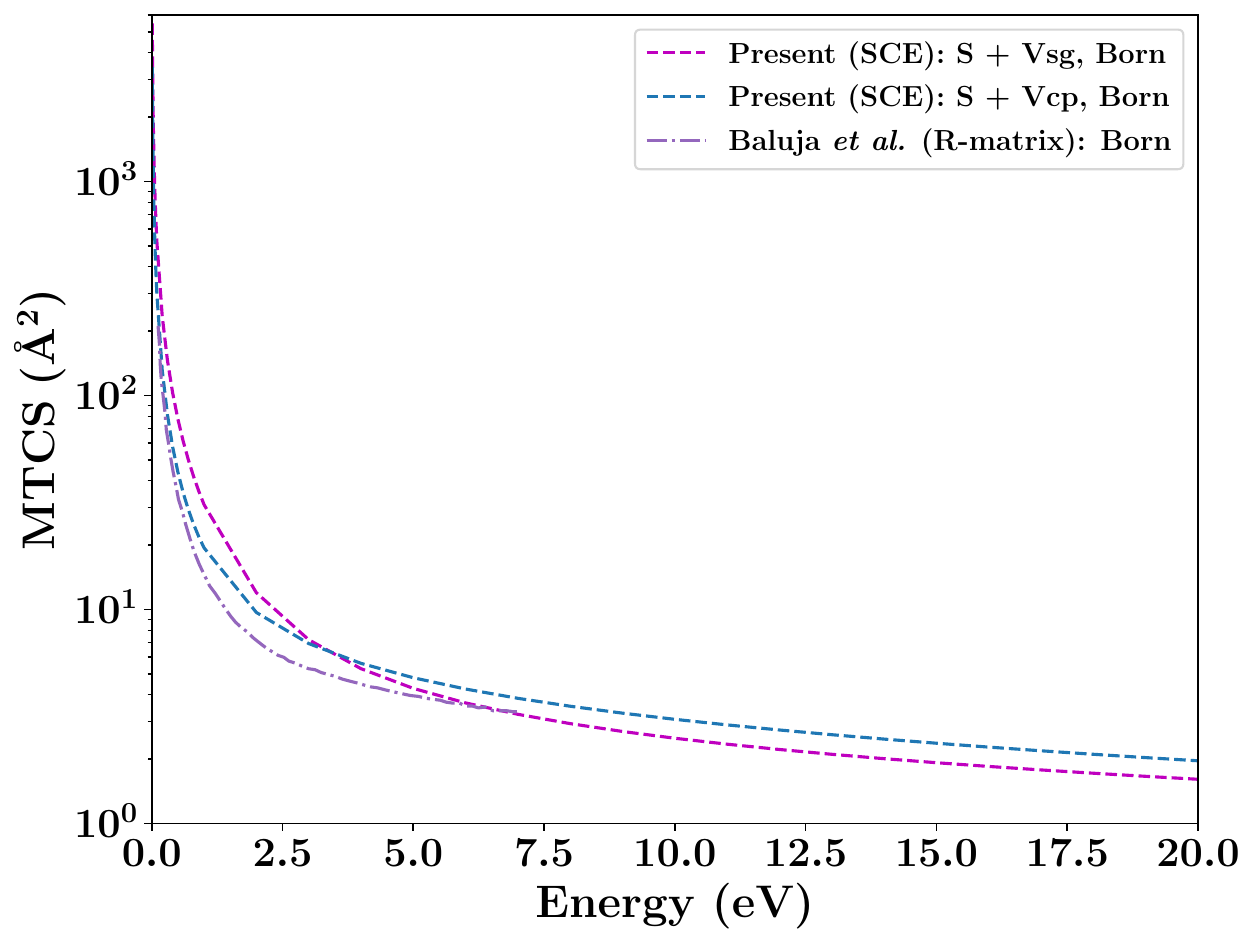}
        \label{fig4:subB}
    }
    \caption{(a) Integral cross-section and (b) momentum transfer cross-section for positron scattering from water molecule. Magenta dashed line: present calculation using $V_{sg}$ correlation. Blue dashed line: present calculation using $V_{cp}$ correlation. Grey dash-dot line: IAM-SCAR+I data of Blanco \textit{et al.} \cite{Blanco_2016}. Brown dash-dot line: Sinha \textit{et al.} \cite{Sinha_2019} SCOP data. Purple dash-dot line: R-matrix data of Baluja \textit{et al.} \cite{Baluja_2007}. Cyan dash-dot line: SMC-SP data of Arretche \textit{et al.} \cite{ARRETCHE2010}. Orange squares: experimental elastic cross-section of Tattersall \textit{et al.} \cite{Tattersall}. Black circles: Makochekanwa \textit{et al.} \cite{Makochekanwa_2009} (Grand total cross-section) - Tattersall \textit{et al.} \cite{Tattersall} (Inelastic+Ionization). Green circles: experimental total cross-section of Kimura \textit{et al.} \cite{kimura1999comparative}. Red triangles: experimental total cross-section of Loreti \textit{et al.} \cite{Loreti_2016}. Yellow stars: experimental total cross-section of Zecca \textit{et al.} \cite{Zecca_2006}.}
    \label{fig:4}
\end{figure*}

The ICS for elastic scattering of positron from water is presented in Fig. \ref{fig4:subA}, up to the incident energy of 20 eV. Water, being a polar molecule, the cross-section will be largely affected by the long-range interaction. This is reflected in the figure by a sharp rise in the cross-sections towards lower energies. Unlike the vast variation in the ICS data for $\rm{V_{sg}}$ and $\rm{V_{cp}}$ correlations in the case of non-polar oxygen molecule, the quantitative difference here is minimal with no qualitative variation. This can be attributed to the Born correction included for the dipolar interaction in both cases, which results in a rapid rise in the cross-section at lower energies. A more intricate comparison between the $\rm{V_{sg}}$ and $\rm{V_{cp}}$ correlations will be seen in the case of DCS, which will be discussed later. Our $\rm{V_{sg}}$ ICS data compares well with the experimental elastic cross-section, obtained as a difference between the grand total cross-section (GTCS) of Makochekanwa \textit{et al.} \cite{Makochekanwa_2009} and total inelastic (electronic excitation+ionization) cross-section of Tattersall \textit{et al.} \cite{Tattersall}. This will be denoted as $ECS_{H_2O}$. The positronium formation cross-section was not subtracted from the GTCS, above the positronium formation threshold of 5.82 eV \cite{Loreti_2016}, as the $\rm{V_{sg}}$ correlation accounts for the virtual positronium formation. The GTCS of Makochekanwa was measured using high-energy resolution instruments and is corrected for the forward scattering angle effects, recommended in the recent experimental cross-section compilation by Brunger \textit{et al.} \cite{Brunger_2017}. Thus, this results in the most reliable and recent corrected elastic cross-section data in the relevant energy range. In the experimental measurements of Tattersall \textit{et al.} \cite{Tattersall}, the available elastic cross-section, which is corrected for forward angle effects, is only for five different energies, with only four data points below 20 eV. These cross-sections are lower in magnitude than our $\rm{V_{sg}}$ and $\rm{V_{cp}}$ ICS and also $ECS_{H_2O}$. 

The TCS of Sueoka \textit{et al.} \cite{sueoka1987total} was corrected by Kimura \textit{et al.} \cite{kimura1999comparative}, however, it has a lower magnitude than the recent corrected cross-sections due to the improvement in the resolution of experimental instruments. The uncorrected TCS data, reported by Zecca \textit{et al.} \cite{Zecca_2006} and the more recent by Loreti \textit{et al.} \cite{Loreti_2016}, are lower than our ICS and the other corrected data, as expected. 

In the theoretical front, the latest ICS data of Blanco \textit{et al.} \cite{Blanco_2016} and Sinha \textit{et al.} \cite{Sinha_2019} are compared with our result. A model potential approach based on the IAM-SCAR method, including the interference effects, is employed by Blanco to calculate the elastic cross-section. The averaged rotational excitation cross-section is also determined within the framework of the first Born approximation. The ICS data of Blanco shown in the figure is the sum of elastic and rotational excitation cross-sections, since our ICS is also rotationally summed. Blanco's cross-section compares well with our $\rm{V_{sg}}$ and $\rm{V_{cp}}$ data, throughout the energy range. Whereas, the SCOP (Spherical complex optical potential) data of Sinha lies quite below, which is expected, as the method does not consider the Born correction for polar molecules. Moving to the presented data of Baluja \textit{et al.} \cite{Baluja_2007} using the R-matrix method and the SMC data of Arretche \textit{et al.} \cite{ARRETCHE2010}, our ICS has a higher magnitude. This can be attributed to the difference in the way of treating the polarization potential, in our case, and these \textit{ab initio} methods. 

Fig. \ref{fig4:subB} illustrates the MTCS for water up to an energy of 20 eV. Our $\rm{V_{sg}}$ and $\rm{V_{cp}}$ data are compared with the MTCS reported by Baluja \textit{et al.} \cite{Baluja_2007} employing the R-matrix method. All the presented results are in good accord with each other, with some differences in magnitude. Given, the only difference between our $\rm{V_{sg}}$ and $\rm{V_{cp}}$ calculations lies in the correlation potential, the difference in magnitude till 20 eV---though small---in the case of ICS and MTCS suggests that some influence of correlation exists even towards higher energies. 

\begin{figure}
    \centering
    \subfigure[]{%
        \includegraphics[width=8cm, height=10cm, keepaspectratio]{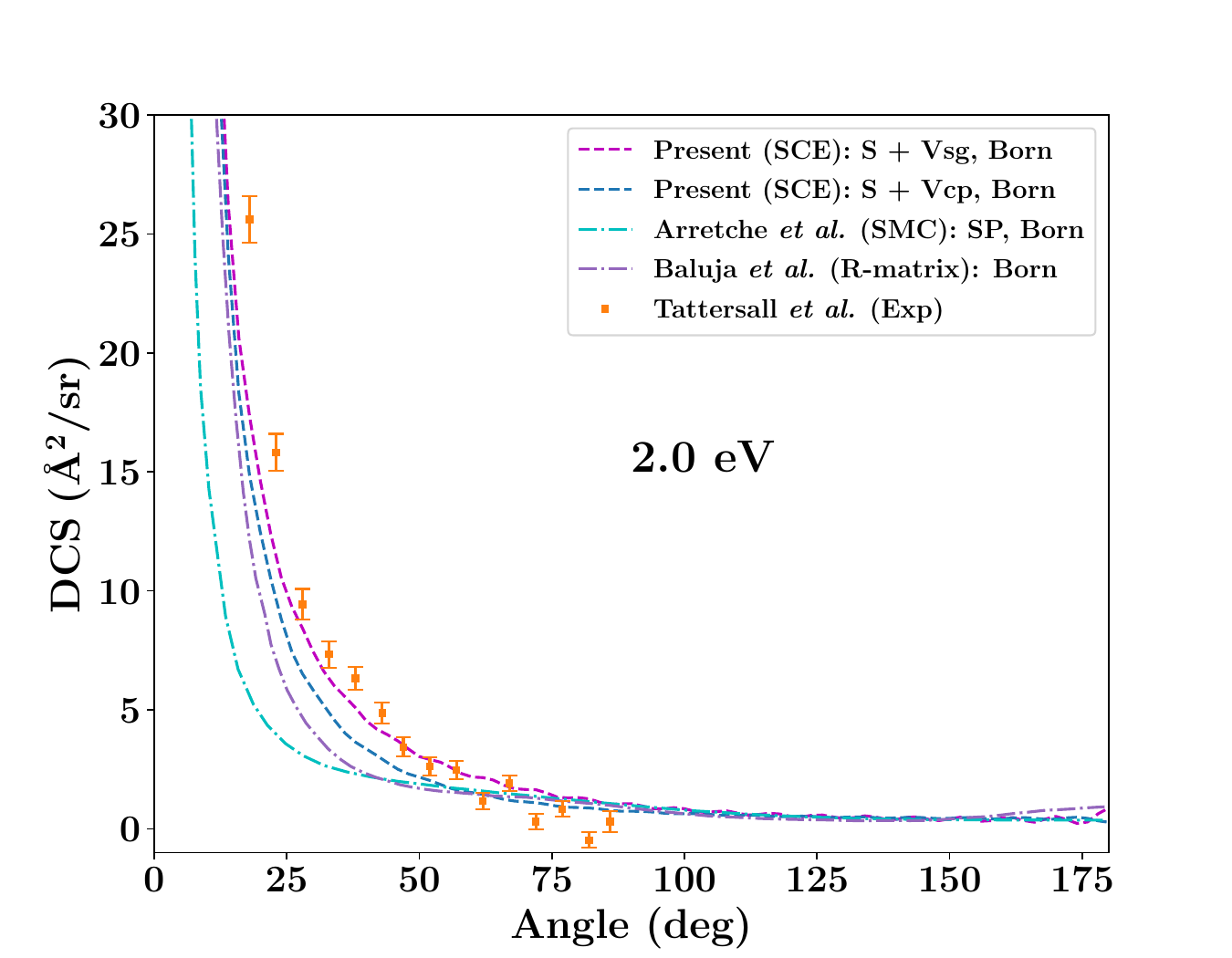}
        \label{fig5:subA}
    }
    \subfigure[]{%
        \includegraphics[width=8cm, height=10cm, keepaspectratio]{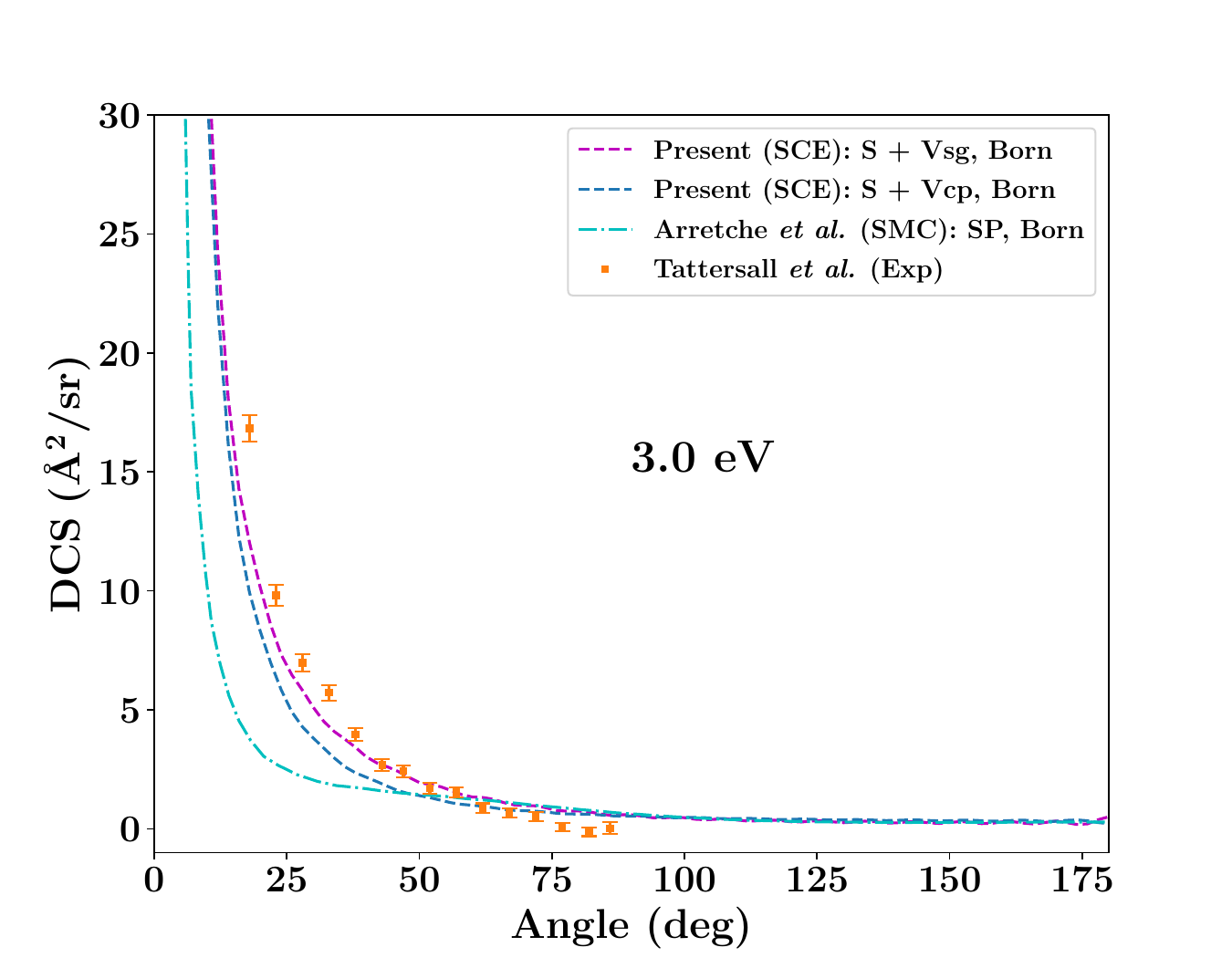}
        \label{fig5:subB}
    }
    \subfigure[]{%
        \includegraphics[width=8cm, height=10cm, keepaspectratio]{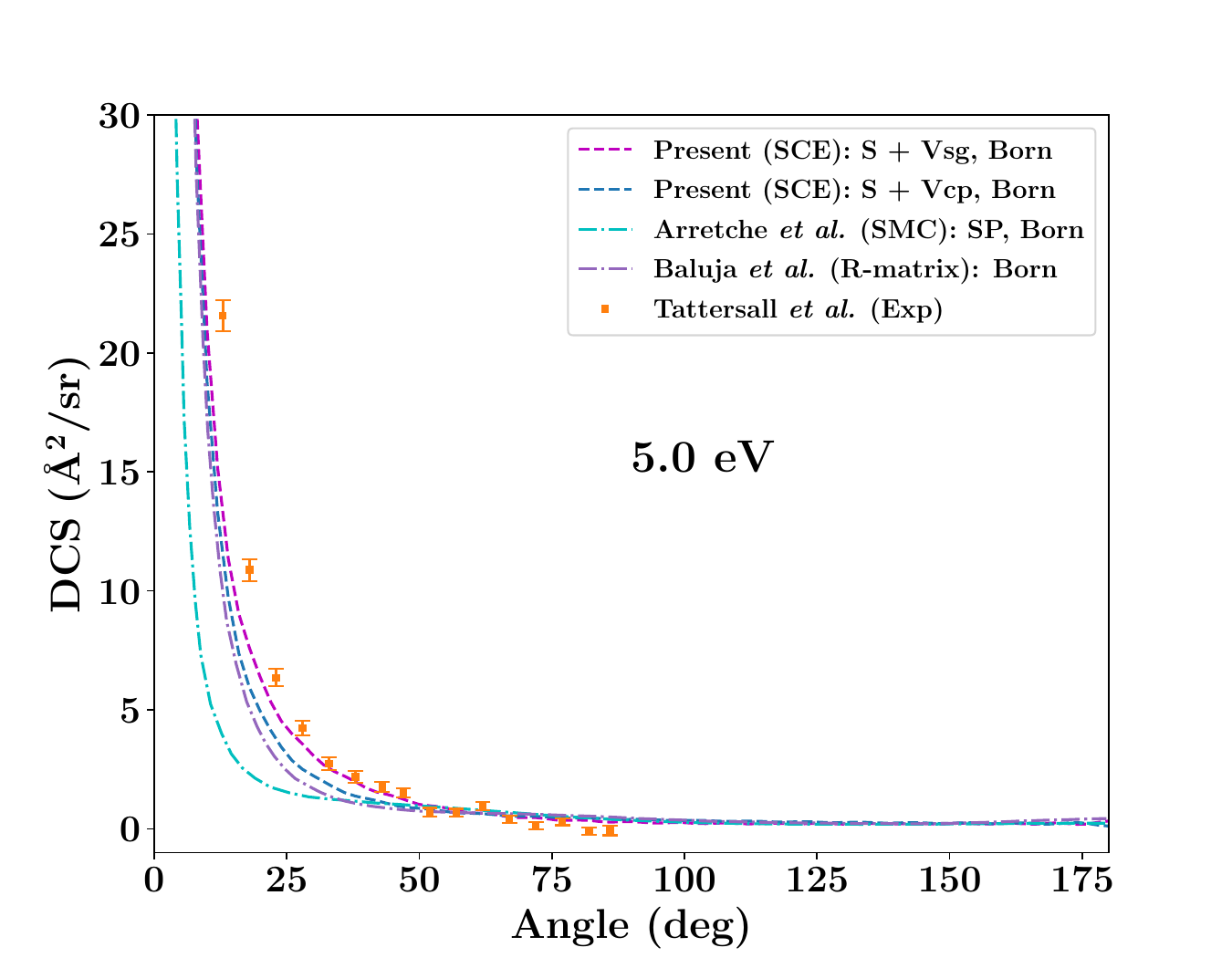}
        \label{fig5:subC}
    }
    % % \subfigure[]{%
    % %     \includegraphics[width=0.45\linewidth]{Figures/8.0ev_DCS_positron_water.pdf}
    % %     \label{fig:subB}
    % }
    \subfigure[]{%
        \includegraphics[width=8cm, height=10cm, keepaspectratio]{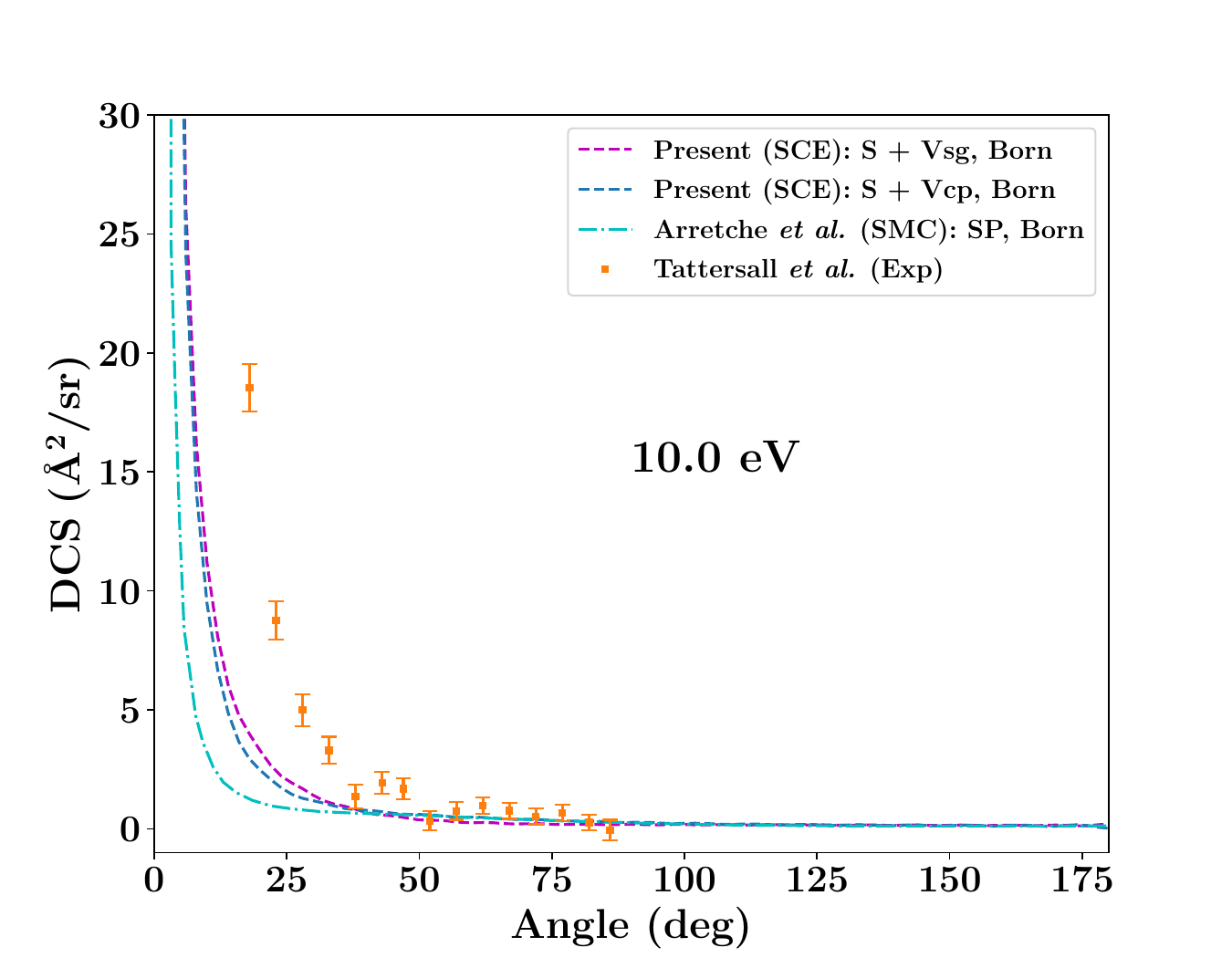}
        \label{fig5:subD}
    }
    % \subfigure[]{%
    %     \includegraphics[width=0.45\linewidth]{Figures/15.0ev_DCS_positron_water.pdf}
    %     \label{fig:subB}
    % }
     \subfigure[]{%
        \includegraphics[width=8cm, height=10cm, keepaspectratio]{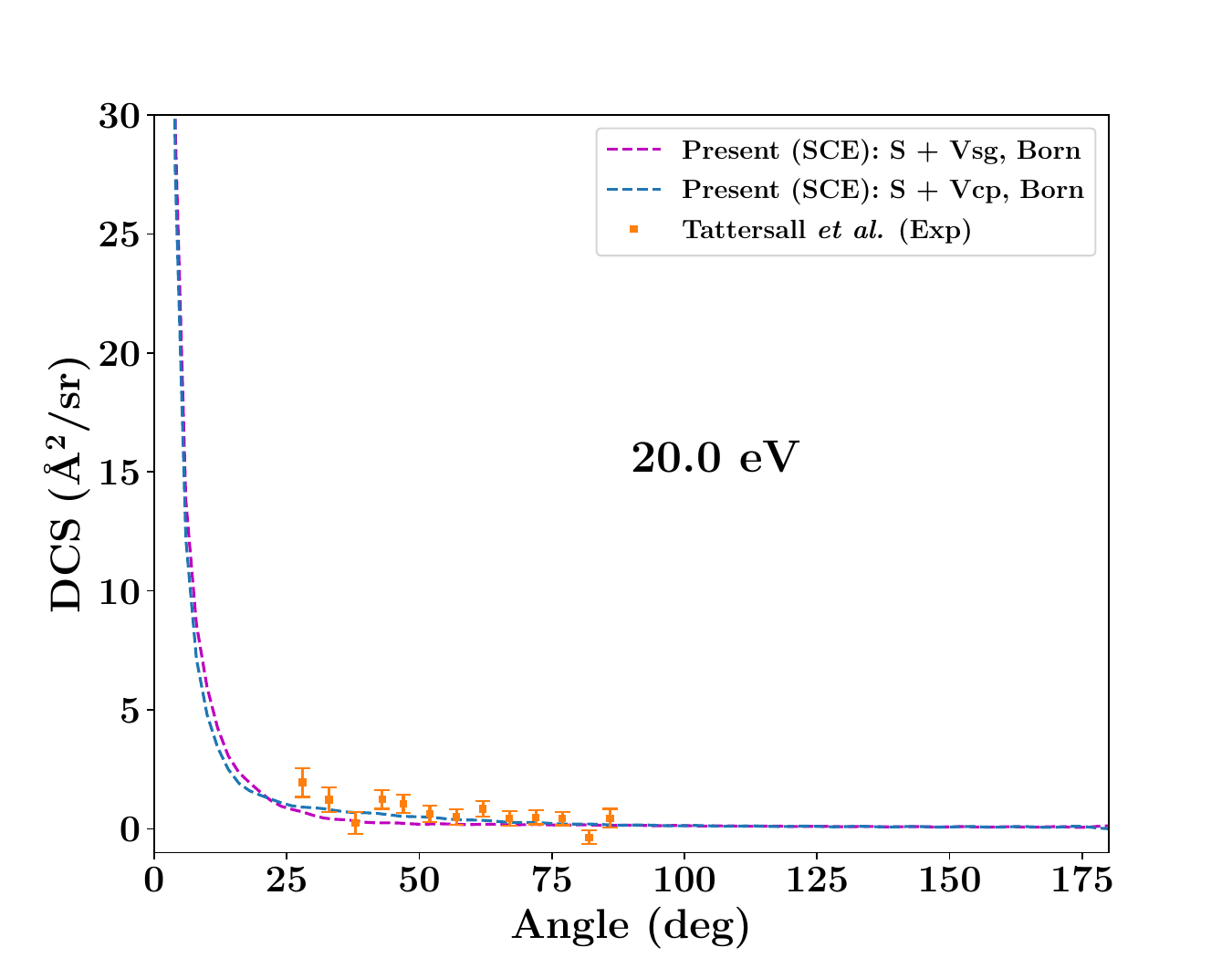}
        \label{fig5:subE}
    }
   
\end{figure}

The DCS for water, presented in Fig. \ref{fig:5}, will better highlight the influence of correlation on the scattering dynamics. All the presented cross-sections are strongly forward-peaked, as expected from a polar molecule. DCS at 2 eV energy (Fig. \ref{fig5:subA}), clearly shows that our $\rm{V_{sg}}$ result compares best with the experimental DCS of Tattersall \textit{et al.} \cite{Tattersall}, below $\rm{75^0}$. Above this angle, both $\rm{V_{sg}}$ and $\rm{V_{cp}}$ data---in addition to the DCS reported by Arretche \textit{et al.} \cite{ARRETCHE2010} and Baluja \textit{et al.} \cite{Baluja_2007}---merge. This is expected as correlation exhibits a crucial role in forward-angle scattering. At 3 eV (Fig. \ref{fig5:subB}) and 5 eV (Fig. \ref{fig5:subC}), the DCS shows a similar trend, with $\rm{V_{sg}}$ data comparing the best with the experimental measurement of Tattersall at forward angles. As the energy increases to 10 eV, the difference in the DCS resulting from $\rm{V_{sg}}$ and $\rm{V_{cp}}$ calculations decreases, and the deviation from the experimental data increases (below $\rm{50^0}$). DCS at 20 eV depicts the diminishing effect of correlation on forward scattering, with $\rm{V_{sg}}$ and $\rm{V_{cp}}$ data having a smaller difference. The experimental data of Tattersall still largely remain in agreement with our result.      

\subsubsection{Formic acid}
\begin{figure}[h]
    \centering
    \subfigure[]{%
        \includegraphics[width=8cm, height=10cm, keepaspectratio]{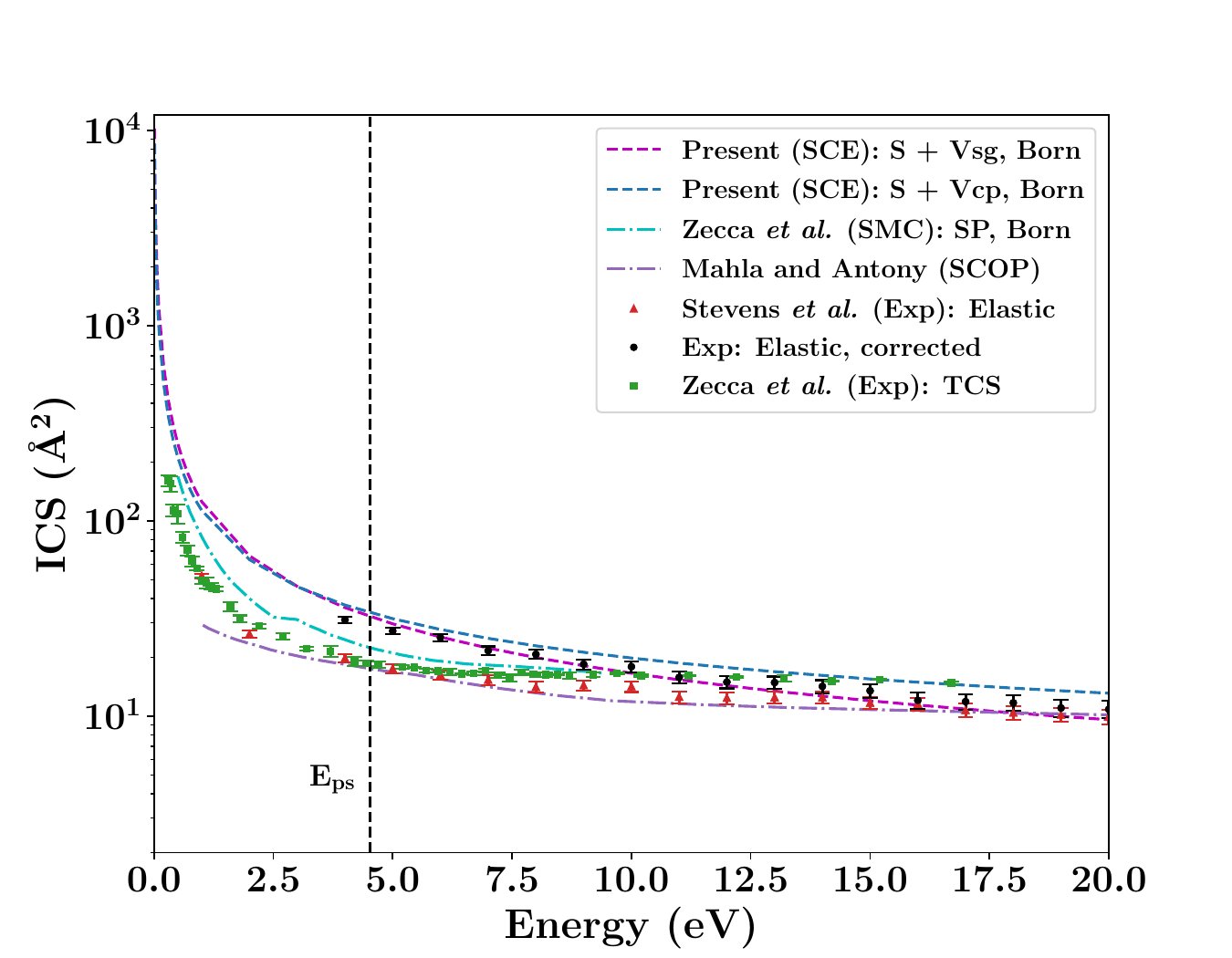}
        \label{fig6:subA}
    }
    \subfigure[]{%
        \includegraphics[width=8cm, height=10cm, keepaspectratio]{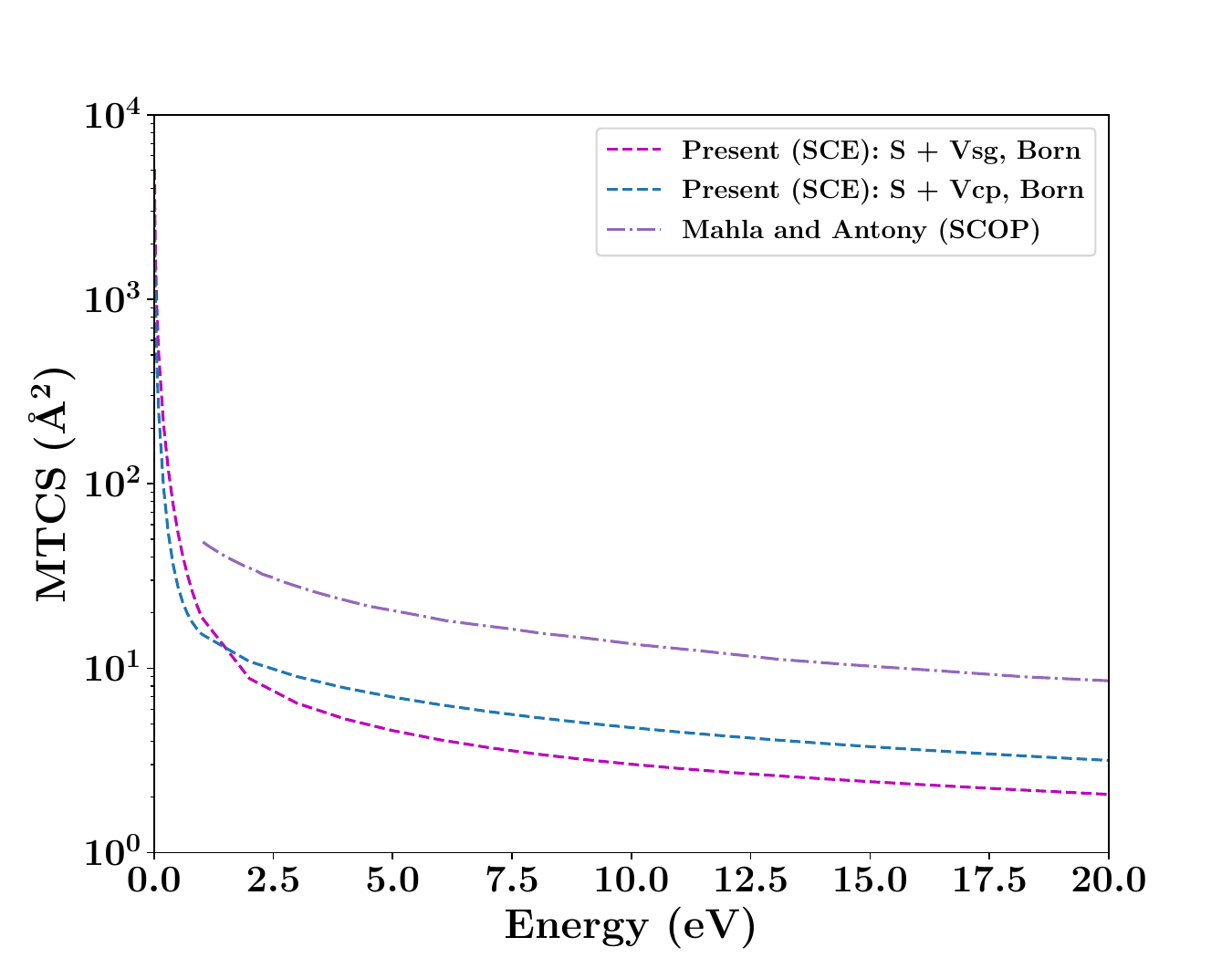}
        \label{fig6:subB}
    }
    \caption{(a) Integral cross-section and (b) momentum transfer cross-section for positron scattering from formic acid. Magenta dashed line: present calculation using $V_{sg}$ correlation. Blue dashed line: present calculation using $V_{cp}$ correlation. Cyan dash-dot line: SMC-SP data of Zecca \textit{et al.} \cite{Zecca_2008}. Purple dash-dot line: Mahla and Antony \cite{mahla2024positron} SCOP data. Red triangles: experimental elastic cross-section of Stevens \textit{et al.} \cite{stevens}. Black circles: Makochekanwa \textit{et al.} \cite{Makochekanwa_2009} (Grand total cross-section) - Stevens \textit{et al.} \cite{stevens} (Inelastic+Ionization). Green squares: experimental total cross-section of Zecca \textit{et al.} \cite{Zecca_2008}.}
    \label{fig:6}
\end{figure}

Fig. \ref{fig6:subA} illustrates the calculated ICS of positron scattering from formic acid, up to the energy of 20 eV. As expected, the cross-section is forward peaked because formic acid has a permanent dipole moment that demands correcting the data within the first Born approximation. Similar to the polar water molecule, the $\rm{V_{sg}}$ and $\rm{V_{cp}}$ ICS are of the same shape with a difference in magnitude. To get good experimental elastic cross-section data, we subtracted the sum of electronic excitation cross-section and the ionization cross-section of Stevens \textit{et al.} \cite{stevens} from the GTCS reported by Makochekanwa \textit{et al.} \cite{Makochekanwa_2009}. This will be referred to as $ECS_{HCOOH}$. The GTCS of Makochekanwa is corrected for the forward angle scattering effects, thereby providing reliable data for comparison. We didn't subtract the positronium formation cross-section (threshold=4.53 eV \cite{stevens}) from the GTCS, as given in \cite{Makochekanwa_2009}, in order to correlate it with the impact of virtual positronium formation on the $\rm{V_{sg}}$ ICS. As we can see, the $ECS_{HCOOH}$ compares very well with our $\rm{V_{sg}}$ ICS, throughout the energy range. This suggests that, even without explicitly including the positronium formation cross-section in our data, the inclusion of the virtual positronium formation in $\rm{V_{sg}}$ correlation somewhat compensates for that. The latest experimental elastic cross-section reported by Stevens \textit{et al.} \cite{stevens}, which is uncorrected for the forward angle scattering effects, also merges with $\rm{V_{sg}}$ ICS, after 12.5 eV.                     

The TCS measured by Zecca \textit{et al.} \cite{Zecca_2008} is lower than our ICS, below 10 eV. The TCS is also lower than their theoretical ICS, reported employing the static plus polarization approximation within the SMC method, with Born correction. One of the major reasons for this is the exclusion of correction for the forward angle scattering effects in the TCS measurement \cite{Zecca_2008}. Our ICS gives a fair comparison with the theoretical ICS of Zecca, at least qualitatively. Other than the theoretical elastic cross-section reported in \cite{Zecca_2008}, Mahla and Antony in their recent work \cite{mahla2024positron}, calculated the ICS for positron-formic acid scattering using the SCOP method. Their ICS mostly lies below almost all the compared theoretical and experimental data, merging with our $\rm{V_{sg}}$ ICS, experimental elastic cross-section of Stevens, and the $ECS_{HCOOH}$ data, roughly above 17 eV. 

Overall, our $\rm{V_{sg}}$ ICS shows a good agreement with the compared experimental data. However, the $\rm{V_{cp}}$ ICS has a comparatively higher magnitude than all the given cross-sections, roughly above 4 eV. Below this energy, the difference in magnitude between $\rm{V_{cp}}$ and $\rm{V_{sg}}$ ICS is minimal due to the dominance of dipolar interaction. As mentioned earlier, the DCS data will provide a more intricate comparison.

Both $\rm{V_{cp}}$ and $\rm{V_{sg}}$ MTCS for formic acid are presented in Fig. \ref{fig6:subB}. They follow a similar trend as the ICS, with $\rm{V_{cp}}$ magnitude exceeding $\rm{V_{sg}}$, roughly above 2 eV. Below that, both the MTCSs are forward-peaked and tend to merge. The SCOP data of Mahla and Antony \cite{mahla2024positron} for MTCS surprisingly lies above our data, unlike the case in ICS.

\begin{figure}
    \centering
    % \subfigure[]{%
    %     \includegraphics[width=0.45\linewidth]{Figures/1ev_DCS_positron_formic_acid.pdf}
    %     \label{fig:subA}
    % }
    \subfigure[]{%
        \includegraphics[width=8cm, height=10cm, keepaspectratio]{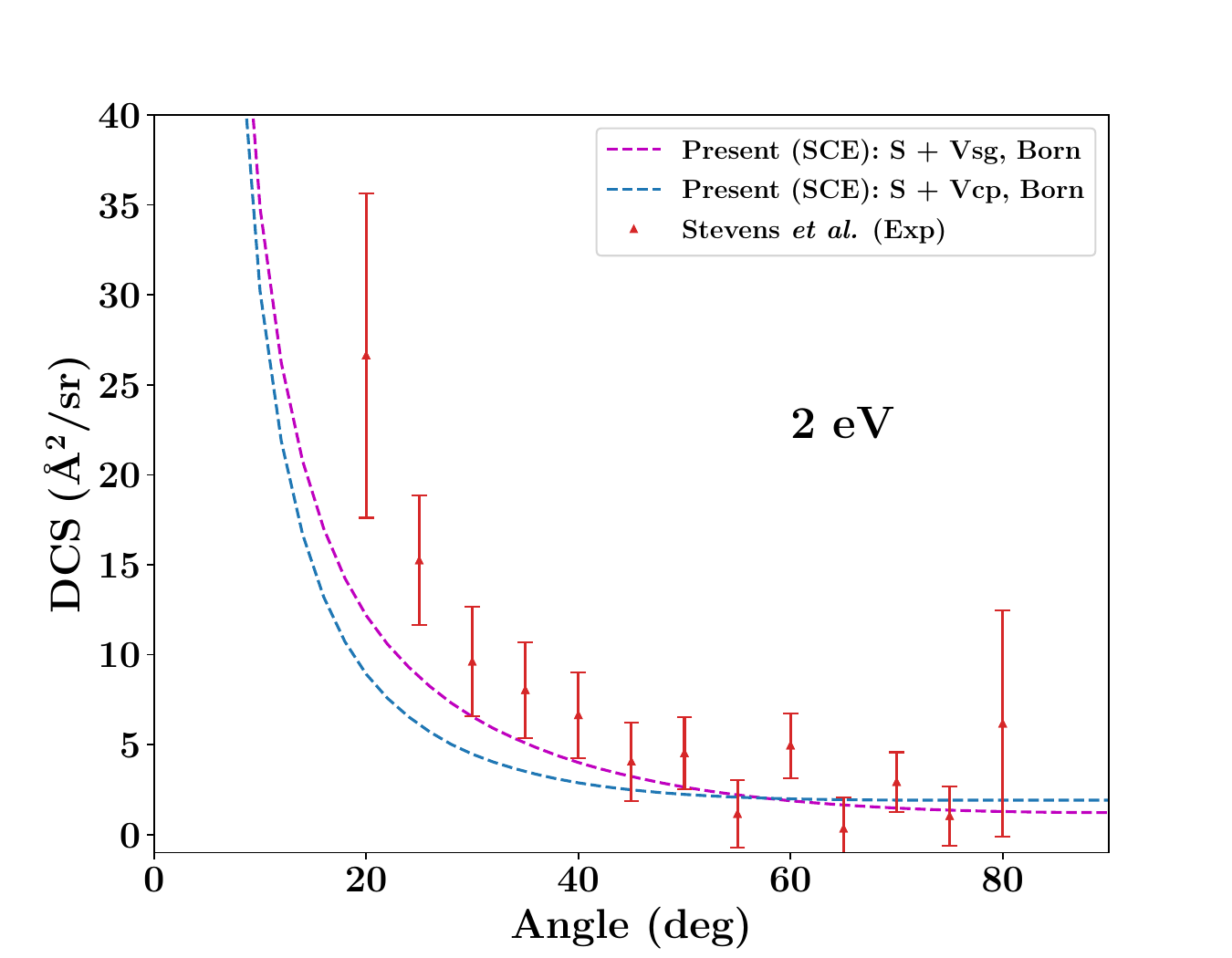}
        \label{fig7:subA}
    }
    \subfigure[]{%
        \includegraphics[width=8cm, height=10cm, keepaspectratio]{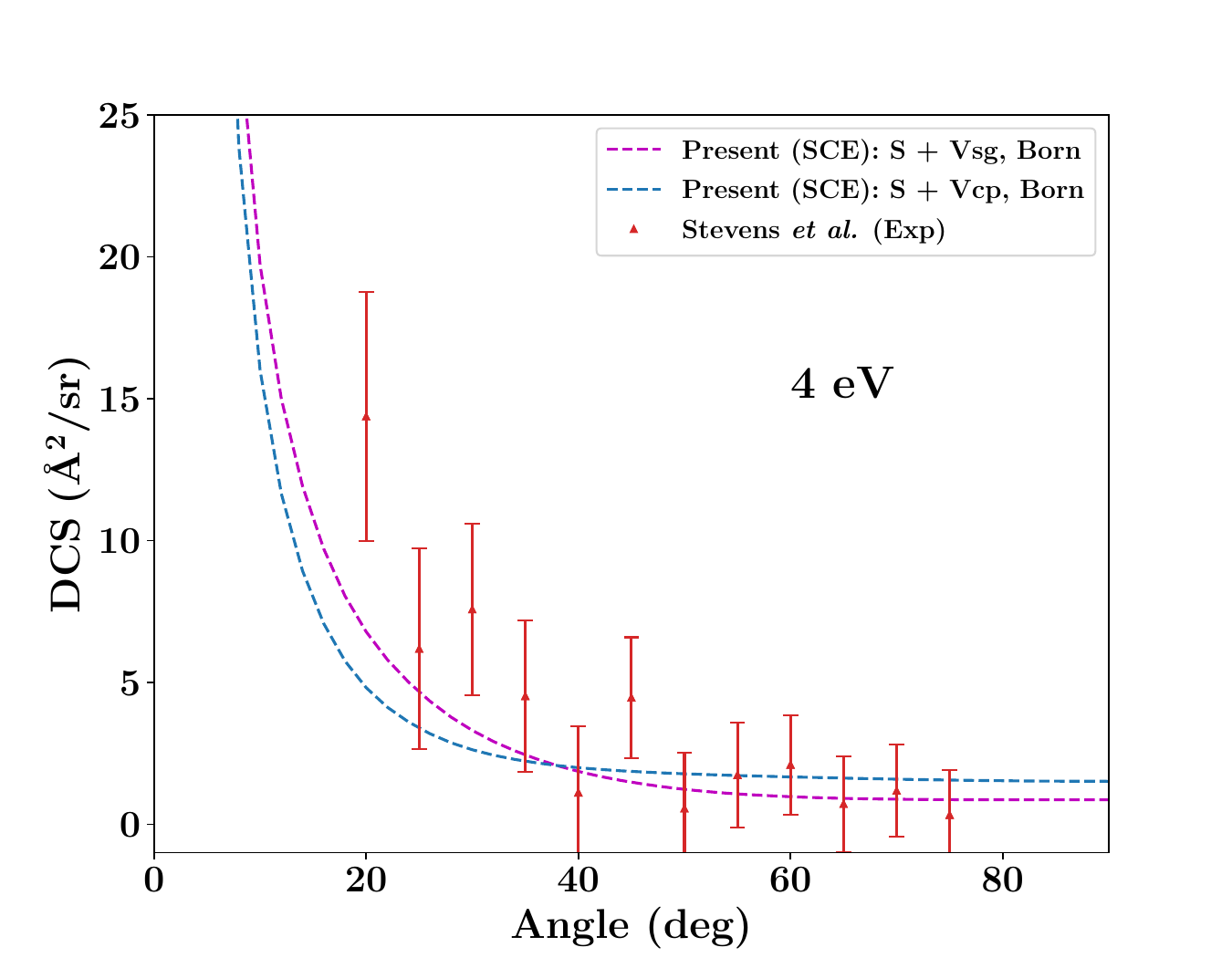}
        \label{fig7:subB}
    }
    \subfigure[]{%
        \includegraphics[width=8cm, height=10cm, keepaspectratio]{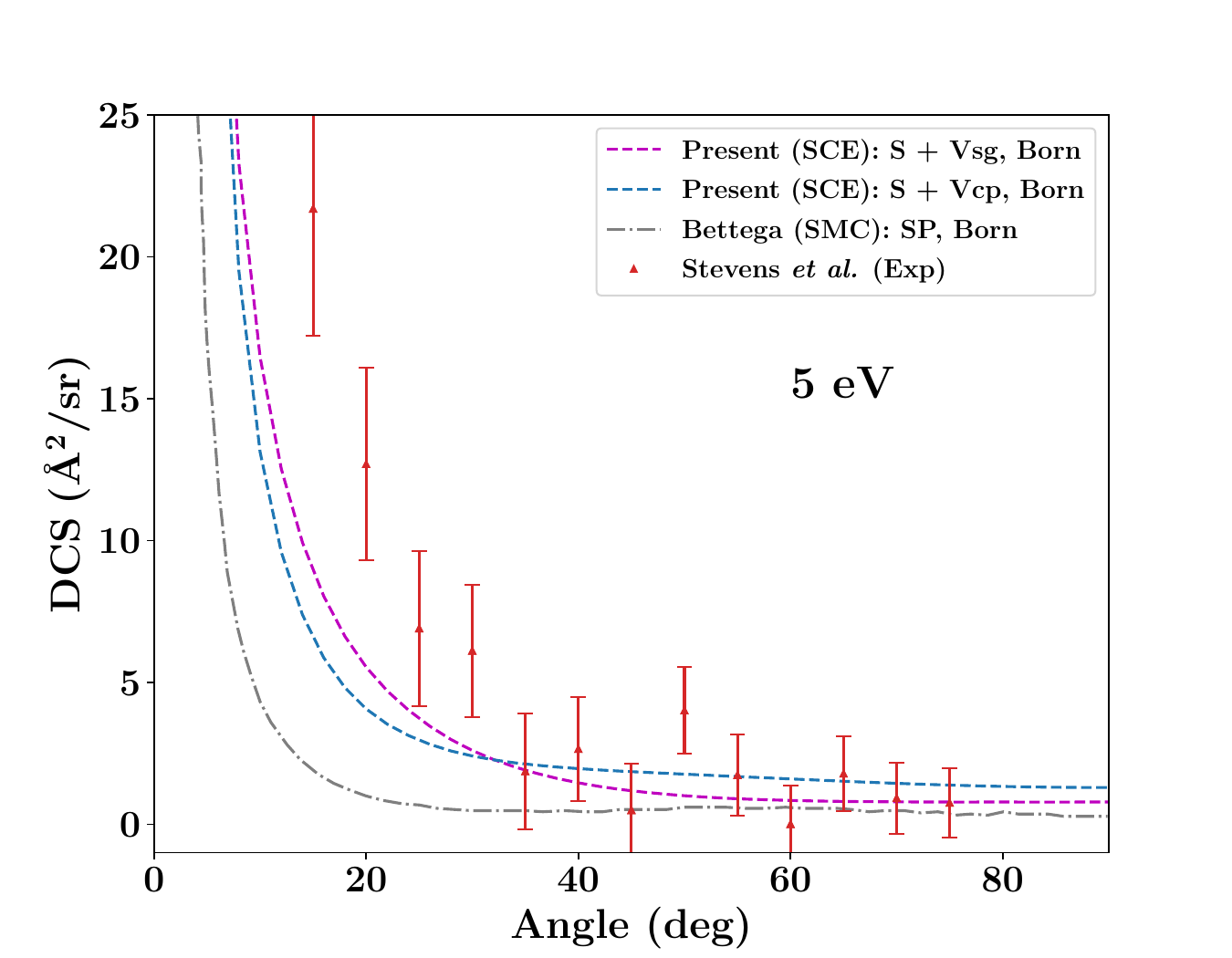}
        \label{fig7:subC}
    }
    \subfigure[]{%
        \includegraphics[width=8cm, height=10cm, keepaspectratio]{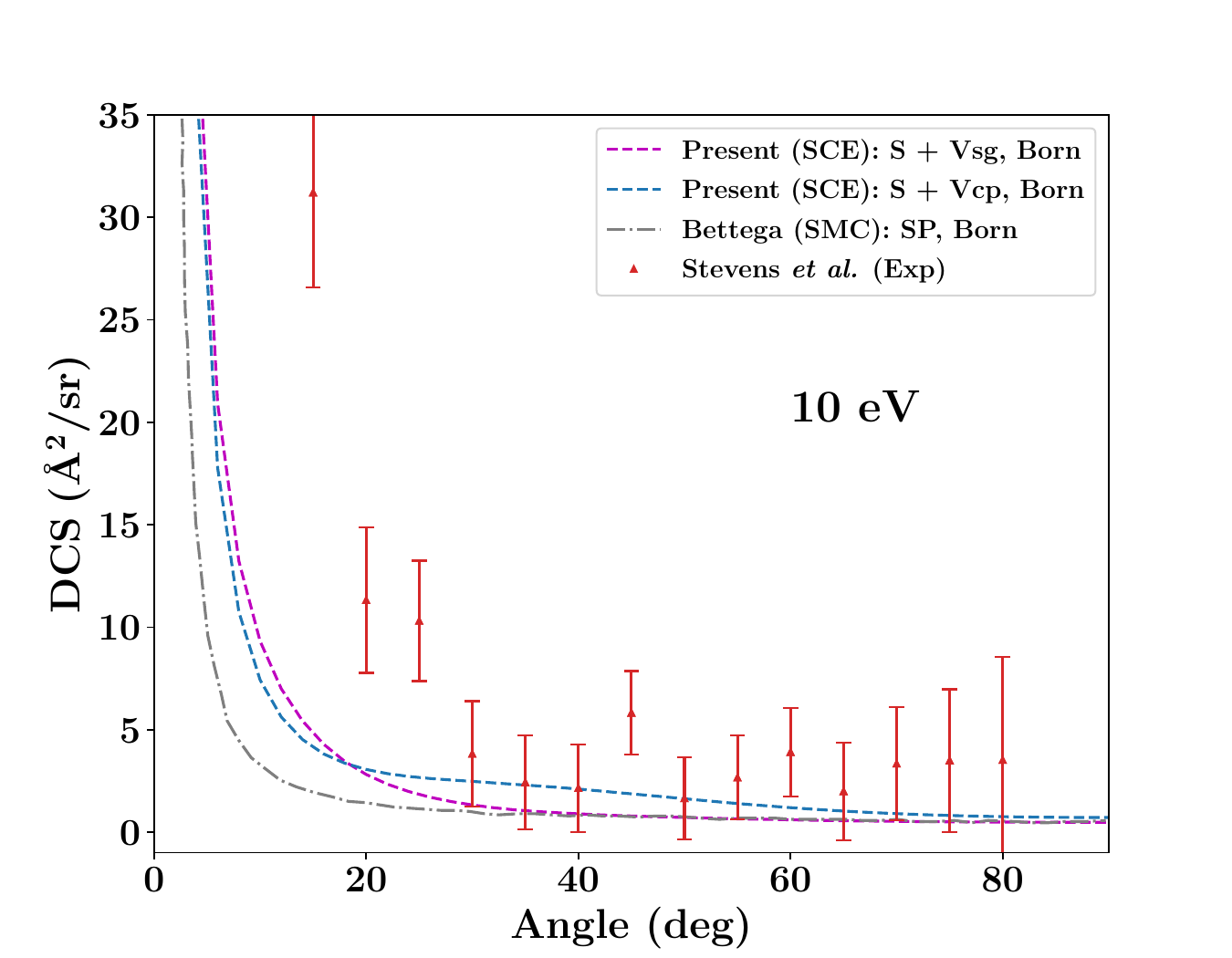}
        \label{fig7:subD}
    }
    % \subfigure[]{%
    %     \includegraphics[width=0.45\linewidth]{Figures/15ev_DCS_positron_formic_acid.pdf}
    %     \label{fig:subB}
    % }
     \subfigure[]{%
        \includegraphics[width=8cm, height=10cm, keepaspectratio]{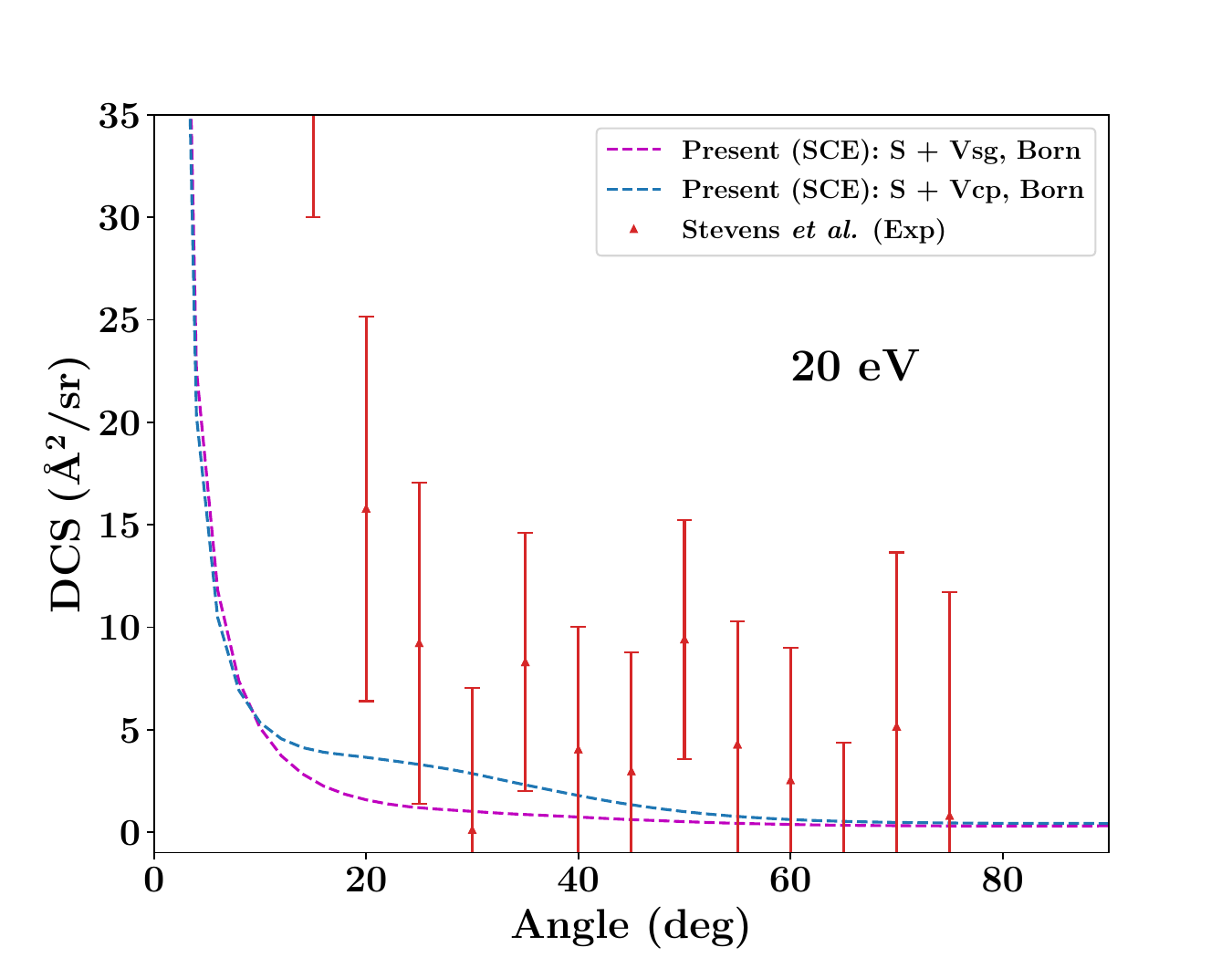}
        \label{fig7:subE}
    }
    
\end{figure}

Despite being an important molecule, formic acid lacks theoretical data for the DCS. In Stevens \textit{et al.} \cite{stevens} work, they have shown the discrepancy between their experimental DCS and the theoretical data of Bettega (SMC method), which they obtained through private communication. Here in Fig. \ref{fig:7}, we present the folded DCS for both $\rm{V_{cp}}$ and $\rm{V_{sg}}$ correlations. We used folded DCS for comparison, as the experimental DCS of Stevens is also folded.
The $\rm{V_{sg}}$ DCS compares quite well with the experimental data of Stevens \textit{et al.} \cite{stevens}, both at 2 eV and 4 eV (Fig. \ref{fig7:subA} and \ref{fig7:subB}). Moving on to the Fig. \ref{fig7:subC}, where we have the theoretical data of Bettega for comparison, the experimental data compare best with our $\rm{V_{sg}}$ DCS, particularly at forward angles. The improvement in the forward angle scattering results suggests the accuracy of the $\rm{V_{sg}}$ model in predicting the correlation between positron and electrons, despite formic acid being a polar molecule where the dominant dipole interaction might mask the correlation effect.

When we move to the higher energies of 10 eV and 20 eV (Fig. \ref{fig7:subD} and \ref{fig7:subE}), our $\rm{V_{sg}}$ DCS drifts away from the experimental data---still in better comparison than the SMC result at 10 eV---for lower scattering angles. The $\rm{V_{sg}}$ and $\rm{V_{cp}}$ data also tend to merge. This is again due to the diminishing effect of correlation-polarization at higher energies, as well as the increasing contribution of higher partial waves, which can result in a different cut-off radius value.   

\subsection{Positron virtual and bound states}

\begin{figure}
    \centering
    \label{fig:8}
    \subfigure[Hydrogen molecule]{%
        \includegraphics[width=8cm, height=10cm, keepaspectratio]{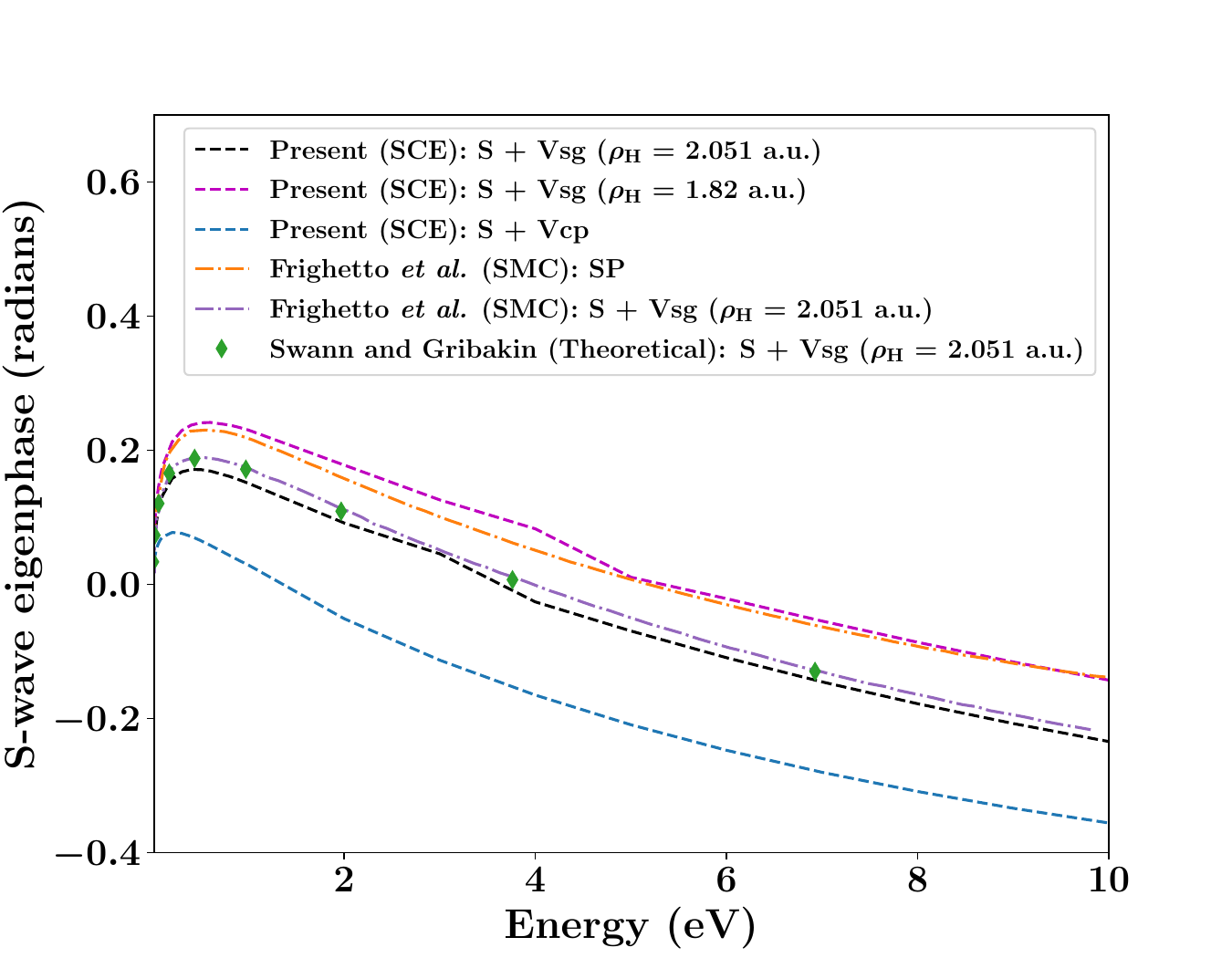}
        \label{fig8:subA}
    }
    \subfigure[Ethylene]{%
        \includegraphics[width=8cm, height=10cm, keepaspectratio]{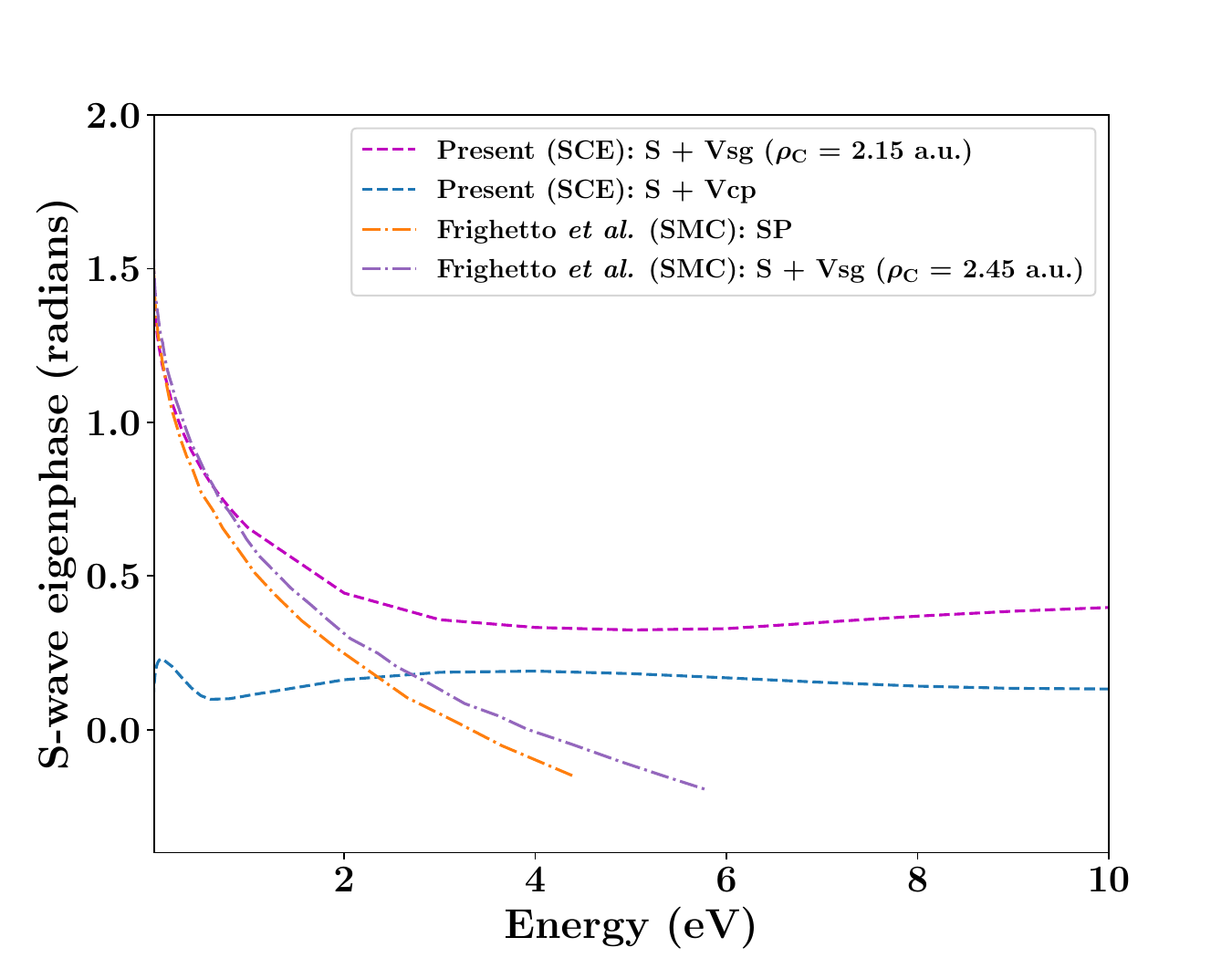}
        \label{fig8:subB}
    }
    \subfigure[Acetylene]{%
        \includegraphics[width=8cm, height=10cm, keepaspectratio]{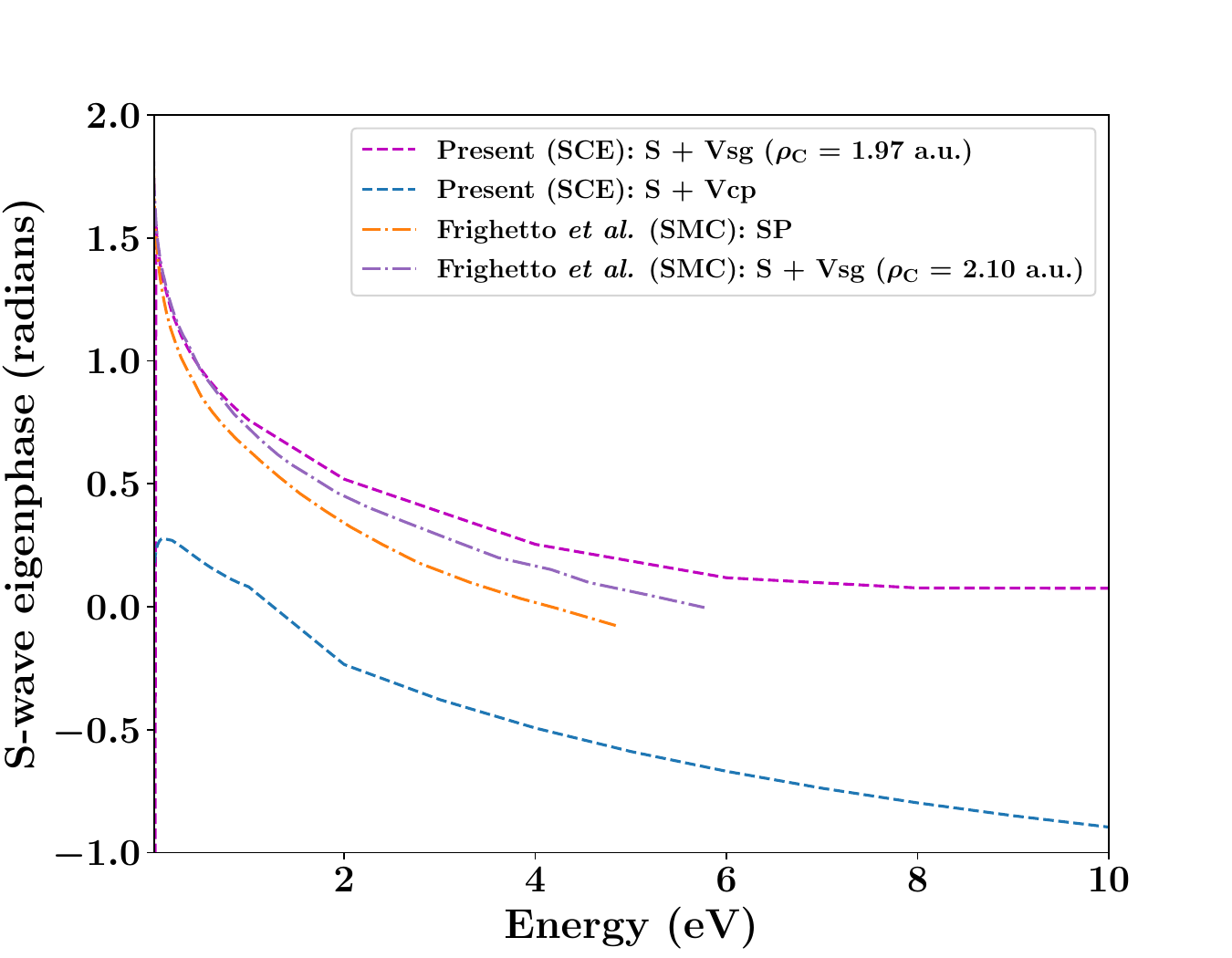}
        \label{fig8:subC}
    }
    \subfigure[Oxygen molecule]{%
        \includegraphics[width=8cm, height=10cm, keepaspectratio]{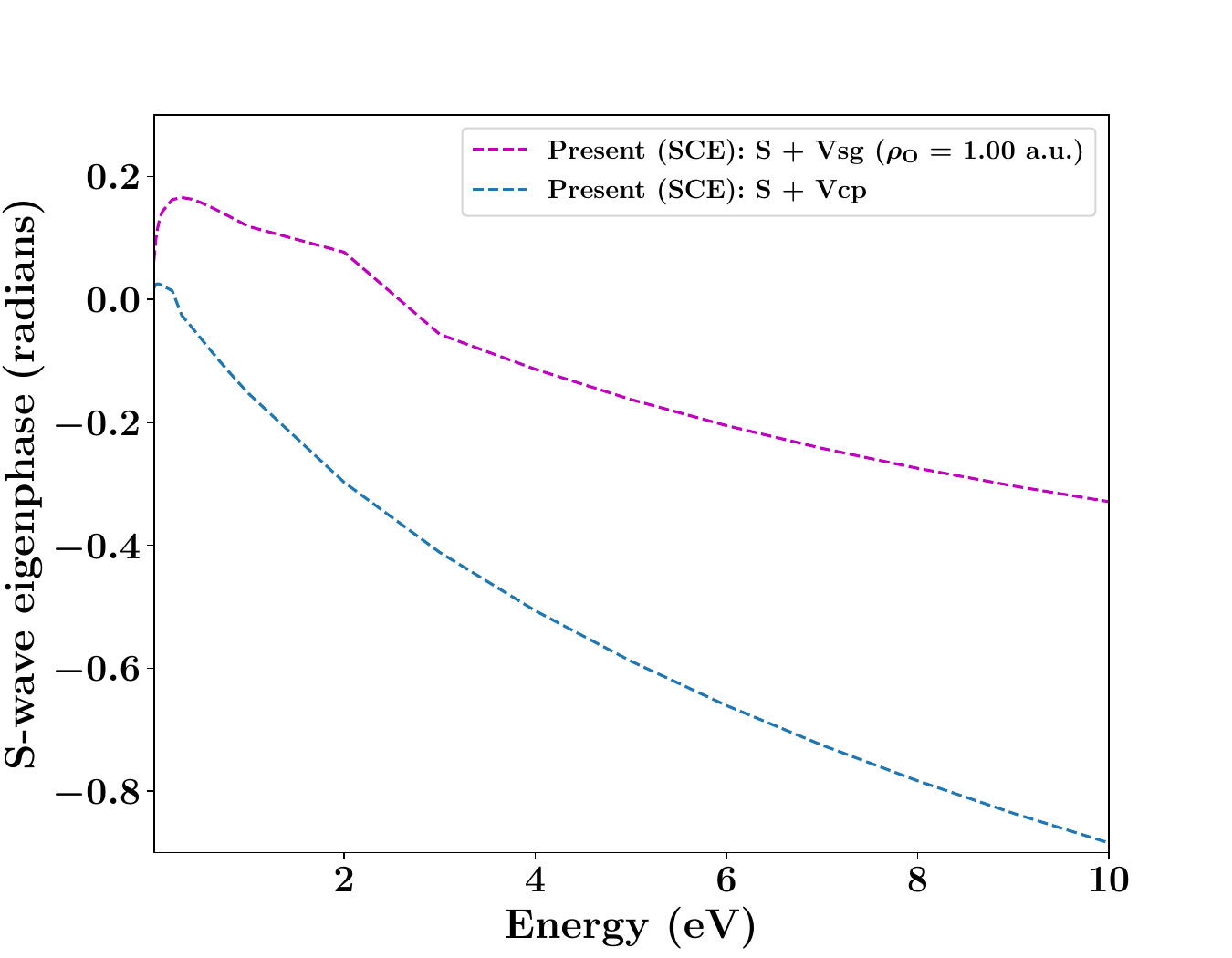}
        \label{fig8:subD}
    }
    \subfigure[Water]{%
        \includegraphics[width=8cm, height=10cm, keepaspectratio]{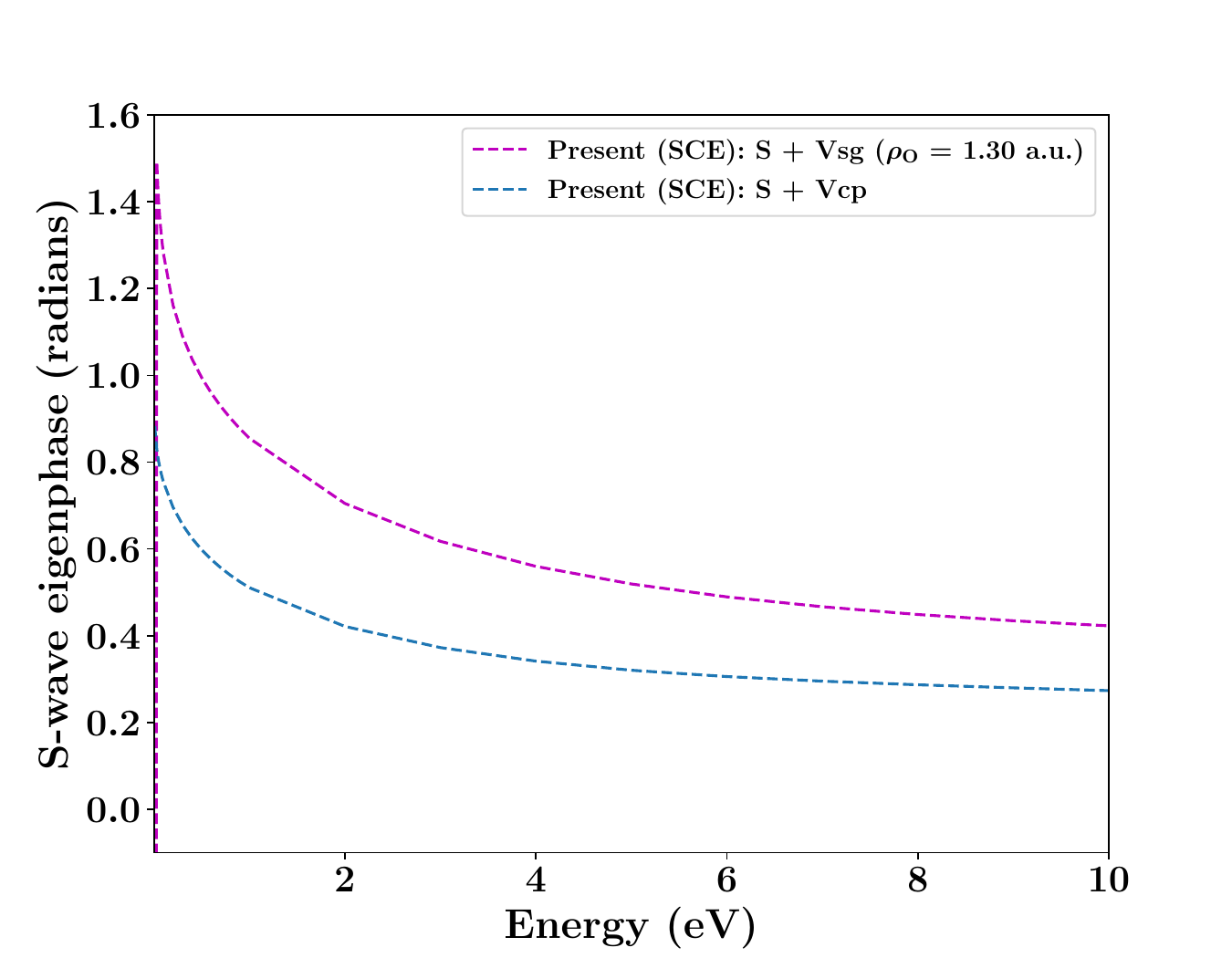}
        \label{fig8:subE}
    }
    \subfigure[Formic acid]{%
        \includegraphics[width=8cm, height=10cm, keepaspectratio]{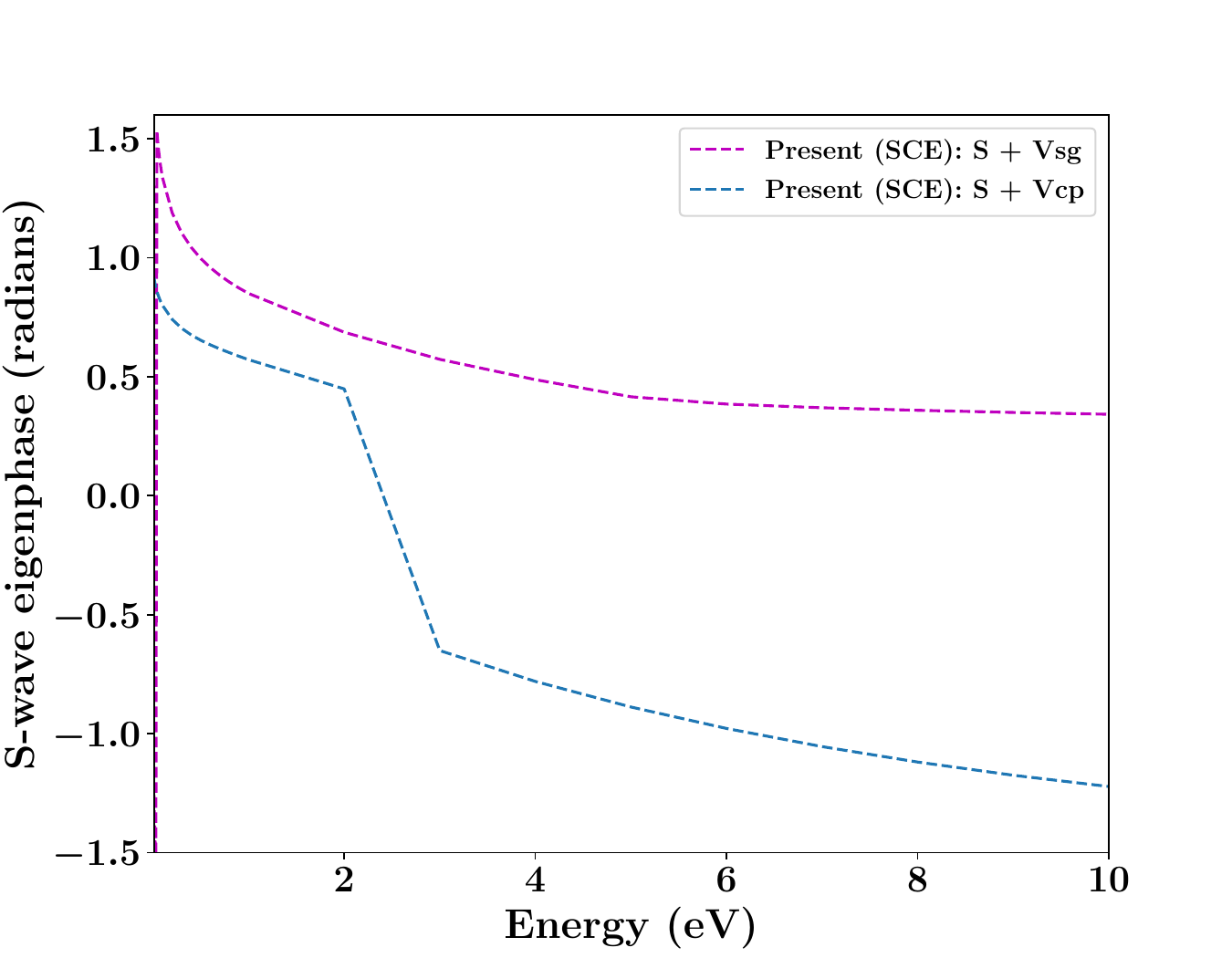}
        \label{fig8:subF}
    }
    
\end{figure}

The formation of positron virtual/bound states with the target molecules can be qualitatively predicted by analyzing the respective s-wave eigenphase of the targets at very low positron energies. A positive slope denotes virtual state formation, whereas a negative slope indicates a bound state. The s-wave eigenphase for the molecules under study is presented in Fig. \ref{fig:8}, calculated for both $\rm{V_{sg}}$ and $\rm{V_{cp}}$ correlations. Both $\rm{V_{sg}}$ and $\rm{V_{cp}}$ eigenphases, as shown in Fig. \ref{fig8:subA}, predict a virtual state for the hydrogen molecule, consistent with the literature. Our $\rm{V_{sg}}$ data with $\rm{\rho_H}$ value of 1.82 a.u. compare well with the best \textit{ab initio} SMC-SP data of Frighetto \textit{et al.} \cite{Frighetto2023Imp}. For a better comparison, we also present the eigenphase data for the $\rm{\rho_H}$ value of 2.051 a.u. Our result is in good agreement with the s-wave eigenphase reported by Frighetto \textit{et al.} \cite{Frighetto2023Imp} (SMC with $\rm{V_{sg}}$ potential and $\rm{\rho_H}=2.051$ a.u.) and the theoretical calculation of Swann and Gribakin \cite{swann2020model} utilizing the same $\rm{V_{sg}}$ correlation and $\rm{\rho_H}$ value. It is evident that $\rm{\rho_H}=1.82$ a.u. gave us a better eigenphase, closer to the \textit{ab initio} result. This again justifies our choice of the $\rm{\rho_H}$ value.  

The scattering length (A in a.u.) and the virtual/bound state energy ($\rm{\epsilon}$ in meV) calculated for the molecules are summarized in Table \ref{Table2}. Negative/positive A and  $\rm{\epsilon}$ values denote virtual/bound state formation. For hydrogen molecule, clearly the $\rm{V_{sg}}$ scattering length is in good agreement with all the reported data in the literature, particularly the ones reported for a bond length of 1.45 $\rm{a_0}$. The present calculation, for $\rm{H_2}$, is also done for a bond length of 1.45 $\rm{a_0}$, which is the mean internuclear distance for its ground vibrational state \cite{kinghorn_1999}.

\begin{table*}[ht]
\centering
% \footnotesize
\begin{tabular}{|c|cc|cc|>{\centering\arraybackslash}p{4cm}>{\centering\arraybackslash}p{4cm}|}
\hline
Target & \multicolumn{2}{c|}{$\rm{S+V_{cp}}$} & \multicolumn{2}{c|}{$\rm{S+V_{sg}}$} & \multicolumn{2}{c|}{\hspace{0.5cm}Literature} \\ 
                & A(a.u.) & $\rm{\epsilon}$(meV) & A(a.u.) & $\rm{\epsilon}$(meV) & A(a.u.) & $\rm{\epsilon}$(meV) \\
\hline
 Hydrogen molecule & $-1.379$ & $-7149.636$ & $-2.824$  & $-1705.735$ & $-2.39$ \cite{swann2020model}, $-2.6$ \cite{Zhang_Positron_2009}, $-2.57 (-2.71)$ \cite{Zhang_Stochastic_2011}, $-2.63$ \cite{Zhang_Scattering_2014}, $-2.49 (-2.65)$ \cite{Zammit_Convergent_2013,Zammit_2017}, $-2.06$ \cite{Zhang_Positron_collisions_2011}, $-2.96 (-3.18)$ \cite{Rawlins_2023}  & -- \\
Ethylene & $-7.196$  & $-262.748$  & $250.751$  & 0.216 & 1300.557 \cite{frighetto2023low} & 0.008 \cite{frighetto2023low}, 0.016 \cite{frighetto2024low}, 4.802 \cite{swann2021effect}, $20\pm8$ \cite{Danielson_2022}, $-17$ \cite{Gribakin} \\
Acetylene & $-7.546$ & $-238.931$ & 78.974 & 2.181 & 140.556 \cite{frighetto2023low} & 0.689 \cite{frighetto2023low}, 0.763 \cite{frighetto2024low}, $-0.8408$ \cite{swann2021effect}, $\ge0$ \cite{swann2021effect}, $-28$ \cite{Gribakin}  \\
Oxygen molecule & $-1.069$ & $-11896.42$ & $-2.674$ & $-1903.164$ & $-5.23$ \cite{pinheiro2023elastic} & --  \\
Water & 200.193 & 0.339 & 71.087 & 2.692 & -- & $-15$ \cite{Gribakin}  \\
Formic acid & 276.357 & 0.178 & 56.344 & 4.286 & -- & -- \\
\hline
\end{tabular}
\caption{Comparison of scattering length (A) and virtual/bound state energy ($\epsilon$) for all the targets using $\rm{V_{cp}}$ and $\rm{V_{sg}}$ correlation models, along with literature values. Negative values denote positron virtual state formation, whereas positive values denote bound state formation. Note: The A values in brackets for the hydrogen molecule are for an internuclear distance of 1.45 $\rm{a_0}$, including our result. Rest are for 1.40 $\rm{a_0}$.}
\label{Table2}
\end{table*}

The negative slope of the $\rm{V_{sg}}$ eigenphase in Fig. \ref{fig8:subB}, for ethylene, suggests a bound state, as is the case with the data reported by Frighetto \textit{et al.} \cite{frighetto2023low,frighetto2024low}. However, the $\rm{V_{cp}}$ eigenphase predicts a virtual state with a positive slope at very low energies. The calculation of the scattering length also suggests a bound state for the $\rm{V_{sg}}$ correlation and a virtual state for $\rm{V_{cp}}$ correlation. Our $\rm{V_{sg}}$ bound state energy of 0.216 meV, though far from the experimental value of $20\pm8$ meV \cite{barnes_2003, Danielson_2022}, is better than that reported by Frighetto \textit{et al.} using both SP approximation (0.008 meV) \cite{frighetto2023low} and $\rm{V_{sg}}$ potential (0.016 meV)  \cite{frighetto2024low}. So far, the model potential calculation of Swann \textit{et al.} \cite{swann2021effect} has given the best approximation of the bound state energy (4.802 meV) for ethylene. The value of $-17$ meV, estimated using a fitting function in \cite{Gribakin}, totally deviates from the experimental prediction.

Although the calculations of Swann \textit{et al.} \cite{swann2021effect} provided a good estimate of positron bound state energy in the case of ethylene, the theoretically predicted virtual state for acetylene ($\rm{\epsilon}=-0.8408$ meV) contradicts their experimental prediction of a bound state ($\rm{\epsilon}\geq0$ meV). Our $\rm{V_{cp}}$ calculation and low-energy eigenphase slope in Fig. \ref{fig8:subC}  also suggest a virtual state. However, the $\rm{V_{sg}}$ calculation estimates a bound state formation of positron with acetylene ($\rm{\epsilon}=2.181$ meV), consistent with the experimental prediction. The SMC calculation of Frighetto \textit{et al.} using SP approximation  \cite{frighetto2023low} and $\rm{V_{sg}}$ potential \cite{frighetto2024low} also predicts a bound state formation, with lower binding energy values. As was the case for ethylene, the fitting function estimation of Gribakin \textit{et al.} \cite{Gribakin} disapproves of a bound state formation for acetylene.

Both $\rm{V_{cp}}$ and $\rm{V_{sg}}$ eigenphases for oxygen molecule in Fig. \ref{fig8:subD} have a positive slope at very-low energies, indicating a virtual state formation. The negative scattering length also depicts the same, with $\rm{V_{sg}}$ scattering length ($-2.674$ a.u.) closer to the value given in literature ($-5.23$ a.u.) \cite{pinheiro2023elastic}. Moving further, the result we obtained for water molecule is quite surprising. Both our eigenphase slope (Fig. \ref{fig8:subE}) and scattering length calculations (Table \ref{Table2}) suggest a bound state formation. This is contrary to the general theoretical and experimental evidence of positron not binding with water \cite{young_and_surko_2008, Gribakin}. However, it is well established that a polar non-rotating molecule having a dipole moment greater than 1.625 Debye is expected to have a bound state with an electron or positron \cite{Assafrão_2010, Tachikawa_2003, Crawford_1967}. Water, indeed, has a dipole moment greater than the critical value and should therefore support positron binding. Therefore, our result---a rather unprecedented one---needs further validation from future experimental and theoretical work. Formic acid too supports positron bound state as evident from Fig. \ref{fig8:subF} and Table \ref{Table2}. To the best of our knowledge, no data are available in the literature on positron binding with formic acid. The sudden rise in eigenphase at low energy, observed for acetylene, water, and formic acid, may depict the formation of bound states similar to resonance in the case of electron scattering. However, it might also be some computational artifact.

\section{\label{sec:level5}Conclusion\protect}
Present work highlights the effect of positron correlation on the scattering cross-sections and the bound/virtual state formation for non-polar as well as polar molecules. We implemented two model correlations in our work: one is the widely used potential by Perdew and Zunger ($V_{cp}$), and other, the recent one by Swann and Gribakin ($V_{sg}$). The latter takes care of the many-body correlations, such as virtual positronium formation, and consists of a free parameter ($\rho_A$). We have optimized the $\rho_A$ values for hydrogen, carbon, as well as oxygen atoms, which will be useful for future calculations involving bigger oxygen-containing molecules.

For non-polar molecules---hydrogen, ethylene, acetylene, and oxygen---the two correlations lead to cross-sections that show both qualitative and quantitative differences. The $V_{sg}$ correlation clearly gives a better result, consistent with the literature. The polar molecules--- water and formic acid---however, result in cross-sections having similar form but differing in magnitude for both $V_{sg}$ and $V_{cp}$ correlations. This is mainly due to the dominating influence of dipole moment on low-energy scattering dynamics. The real test of the correlation is given by the differential cross-section, where $V_{sg}$ outperforms $V_{cp}$ even for polar molecules. 

The positron virtual/bound state formation, estimated using the $V_{sg}$ and $V_{cp}$ eigenphases, gave some interesting results. While the $V_{sg}$ eigenphase predicted a bound state for ethylene and acetylene, consistent with the literature, the $V_{cp}$ eigenphase estimated a virtual state. For hydrogen molecule, the calculated scattering length compares very well with the literature for $V_{sg}$ correlation. The $V_{sg}$ scattering length for oxygen molecule is also in a decent comparison with the literature. In the case of water molecule, both the $V_{sg}$ and $V_{cp}$  scattering lengths depict a bound state formation. Although this is consistent with the critical dipole moment prediction, it is inconsistent with the literature predicting a virtual state for water molecule. For formic acid too, we have predicted a positron bound state. There is no data in the literature to back this. Thus, further experimental as well as theoretical work is needed to testify our results for water and formic acid.

Lastly, we can say that the single-center expansion method, along with the $V_{sg}$ correlation, results in good cross-sections and virtual/bound state prediction for positron scattering from both non-polar and polar molecules. Thus, this approach---simpler and computationally cheaper than the \textit{ab initio} methods---can be implemented to study positron interaction with larger molecules. An even more intricate result can be achieved if we take into account the higher-order polarizabilities in the polarization potential, as well as the anisotropy involved. Also, different cut-off radii for higher $l$ values will give a better result towards higher energies. 

\section*{Acknowledgments}
DG acknowledges the Science and Engineering Research Board (SERB), Department of Science and Technology (DST), Government of India (Grant No. SRG/2022/000394) for providing a computing facility. This work was financially supported by Vellore Institute of Technology (VIT), Vellore under the Faculty Seed Grant (RGEMS) (Sanction Order No.: SG20250016)

\clearpage
\captionsetup{justification=centering}
\captionof{figure}{Integral cross-section of molecules to fix the cut-off radius ($\rho_A$) value for different atoms. Magenta and black dashed line: present calculation using $V_{sg}$ correlation. Blue dashed line: present calculation using $V_{cp}$ correlation. Orange dash-dot line: Frighetto \textit{et al.} \cite{frighetto2024low, frighetto2023low} SMC-SP data. Purple dash-dot line: Frighetto \textit(\textit{et al.}) \cite{Frighetto2023Imp, frighetto2024low} SMC data using $V_{sg}$ correlation. Green dash-dot line: MCF-PD data of Tenfen \textit{et al.} \cite{TENFEN2022}. Black circles: Makochekanwa \textit{et al.} \cite{Makochekanwa_2009} (GTCS) $-$ Tattersall \textit{et al.} \cite{Tattersall} (Inelastic+Ionization).}
 \label{fig:1}

\clearpage
\captionsetup{justification=centering}
\captionof{figure}{Differential cross-section for positron scattering from water molecule. Magenta dashed line: present calculation using $V_{sg}$ correlation. Blue dashed line: present calculation using $V_{cp}$ correlation. Cyan dash-dot line: SMC-SP data of Arretche \textit{et al.} \cite{ARRETCHE2010}. Purple dash-dot line: R-matrix data of Baluja \textit{et al.} \cite{Baluja_2007}. Orange squares: experimental elastic cross-section of Tattersall \textit{et al.} \cite{Tattersall}.}
 \label{fig:5}

\clearpage
\captionsetup{justification=centering}
\captionof{figure}{Differential cross-section for positron scattering from formic acid. Magenta dashed line: present calculation using $V_{sg}$ correlation. Blue dashed line: present calculation using $V_{cp}$ correlation. Grey dash-dot line: SMC data of Bettega taken from \cite{stevens}. Red triangles: experimental data of Stevens \textit{et al.} \cite{stevens}.}
 \label{fig:7}
 
\clearpage
\captionsetup{justification=centering}
\captionof{figure}{S-wave eigenphase of the target molecules. Magenta and black dashed line: present calculation using $V_{sg}$ correlation. Blue dashed line: present calculation using $V_{cp}$ correlation. Orange dash-dot line: Frighetto \textit{et al.} \cite{frighetto2024low, frighetto2023low, Frighetto2023Imp} SMC-SP data. Purple dash-dot line: Frighetto \textit{et al.} \cite{Frighetto2023Imp, frighetto2024low} SMC data using $V_{sg}$ correlation. Green diamonds: Theoretical data of Swann and Gribakin \cite{swann2020model}.}
 \label{fig:8}

\clearpage
\bibliography{apssamp.bib}% Produces the bibliography via BibTeX.

\end{document}